\documentclass[structabstract]{aa}  
\usepackage{graphicx}
\usepackage{txfonts}
\usepackage{natbib}
\usepackage{color}
\newcommand{\comment}[1]{}
\newcommand{\vel}{\upsilon}
\begin{document}
   \title{Three-dimensional simulations of near-surface convection in main-sequence stars}
   \subtitle{I. Overall structure}

   \author{B. Beeck
          \inst{1,2}
          \and
          R.~H. Cameron\inst{1}
	  \and
	  A. Reiners\inst{2}
     	  \and
	  M. Sch\"ussler\inst{1}
          }

   \institute{Max-Planck-Institut f\"ur Sonnensystemforschung,
              Max-Planck-Stra\ss e 2, 37191 Katlenburg-Lindau, Germany\\
         \and
              Institut f\"ur Astrophysik, 
	      Universit\"at G\"ottingen, 
	      Friedrich-Hund-Platz 1,
	      37077 G\"ottingen, Germany\\
             }

   \date{Received: 22 February, 2013 ; accepted: 16 August 2013}
   \titlerunning{3D simulations of stellar surface layers}

\abstract
{The near-surface layers of cool main-sequence stars are structured by convective flows, which are overshooting into the atmosphere. The flows and the associated spatio-temporal variations of density and temperature affect spectral line profiles and thus have an impact on estimates of stellar properties such as effective temperature, gravitational acceleration, and abundances.}
{We aim at identifying distinctive properties of the thermodynamic structure of the atmospheres of different stars and understand their causes.}
{We ran comprehensive 3D radiation hydrodynamics simulations of the near-surface layers of six simulated stars of spectral type F3V to M2V with the \texttt{MURaM} code. We carry out a systematic parameter study of the mean stratifications, flow structures, and the energy flux in these stars.}
{We find monotonic trends along the lower main sequence in granule size, flow velocity, and intensity contrast. The convection in the M-star models differs substantially from that of the hotter stars, mainly owing to the more gradual transition from convective to radiative energy transport.}
{While the basic mechanisms driving surface convection in cool stars are the same, the properties of the convection vary along the main sequence. Apart from monotonic trends in rms velocity, intensity contrast, granule size, etc., there is a transition between ``naked'' and ``hidden'' granulation around spectral type K5V caused by the (highly non-linear) temperature dependence of the opacity. These variations have to be taken into account when stellar parameters are derived from spectra.}


   \maketitle
%

\section{Introduction}
\begin{table*}
\caption{Stellar parameters, bolometric intensities, and rms velocities.}\label{tab:values}
\centering
\begin{tabular}{cr@{.}lr@{.}l@{$\,\pm\,$}r@{.}lr@{.}lr@{.}lr@{.}lr@{.}l}\hline\hline
{ Simulation} & \multicolumn{2}{c}{$\log g$$^{\mathrm{a}}$} & \multicolumn{4}{c}{$T_{\mathrm{eff}}\,[\mathrm{K}]$} &\multicolumn{2}{c}{$\langle I \rangle/ \langle I \rangle_{\mathrm{G2V}}\,^{\mathrm{b}}$} & \multicolumn{2}{c}{$\sigma_I\,^{\mathrm{c}}$ [\%]} & \multicolumn{4}{c}{$\vel_{z, \mathrm{rms}}(z)\,[\mathrm{km\,s^{-1}}]$}\\ 
 &\multicolumn{2}{c}{} & \multicolumn{4}{c}{} &\multicolumn{2}{c}{} & \multicolumn{2}{c}{} & \multicolumn{2}{c}{$z=z_0\,^{\mathrm{d}}$} & \multicolumn{2}{c}{$z=z_2\,^{\mathrm{e}}$}\\\hline
{ F3V} & 4&301 & 6893&2 & 6&4 & 1&98 & 20&3 & 5&1 & 3&5 \\
{ G2V} & 4&438 & 5764&4 & 7&4 & 1&00 & 15&7 & 2&0 & 1&2 \\
{ K0V} & 4&609 & 4855&6 & 5&5 & 0&500 & 8&04 & 1&1 & 0&72 \\
{ K5V} & 4&699 & 4367&9 & 2&0 & 0&322 & 6&90 & 0&99 & 0&57 \\
{ M0V} & 4&826 & 3904&8 & 0&8 & 0&195 & 3&55 & 0&72 & 0&42 \\
{ M2V} & 4&826 & 3688&6 & 1&4 & 0&153 & 2&17 & 0&59 & 0&57 \\\hline
\end{tabular}
\begin{list}{}{}
\item[$^{\mathrm{a}}$] $g$ is the gravitational acceleration at the stellar surface in cgs units.
\item[$^{\mathrm{b}}$] temporal and spatial mean of the bolometric intensity normalised to the solar value
\item[$^{\mathrm{c}}$] bolometric intensity contrast (temporal mean)
\item[$^{\mathrm{d}}$] $z_0:=\langle z(\tau_{\mathrm{R}}=1)\rangle$
\item[$^{\mathrm{e}}$] $z_2$ is defined as the depth where $\langle p(z_2) \rangle=100\,\langle p(z_0)\rangle$
\end{list}
\end{table*}
\begin{table*}
\centering
\caption{Box sizes and grid resolutions.}\label{tab:H_p}
\begin{tabular}{lr@{}lr@{}lr@{}lr@{}lr@{}lr@{}l}\hline\hline
Simulation & \multicolumn{2}{c}{ F3V} & \multicolumn{2}{c}{ G2V} &  \multicolumn{2}{c}{ K0V} &  \multicolumn{2}{c}{ K5V} &  \multicolumn{2}{c}{ M0V} &  \multicolumn{2}{c}{ M2V} \\\hline
Box height [Mm] & 9& & 3& & 1&.8 & 1&.5 & 0&.9 & 0&.8 \\ 
~~~~~above $z_0$$^{\mathrm{a}}$ & 1&.57 & 0&.95 & 0&.48 & 0&.41 & 0&.25 & 0&.21\\
~~~~~below $z_0$ & 7&.43 & 2&.05 & 1&.32 & 1&.09 & 0&.65 & 0&.59\\\hline
\# of pressure scale heights & 13&.1 & 14&.0 & 13&.2 & 15&.3 & 14&.8
 & 14.5\\
~~~~~above $z_0$ & 6&.9 & 8&.6 & 7&.1 & 9&.1 & 8&.4 & 7&.8 \\
~~~~~below $z_0$ & 6&.2 & 5&.4 & 6&.2 & 6&.4 & 6&.4 & 6&.8 \\\hline
$H_p$ at $z_0$ [km] & 500& & 200& & 90& & 65& & 38& & 35& \\\hline
$\Delta z\,^{\mathrm{b}}$ [km] & 11&.25 & 10& & 6& & 5& & 4& & 3&.2\\
$\min(H_p$)/$\Delta z$ & 18&.1 & 10&.0 & 9&.57 & 7&.87 & 6&.53 & 7&.19\\\hline
Horizontal box size [Mm] & 30& & 9& & 6& & 4& & 2&.5 & 1&.56\\\hline
$\Delta x,\Delta y\,^{\mathrm{c}}$ [km] & 58&.6 & 17&.6 & 11&.7 & 7&.81 & 4&.88 & 3&.05\\
$\Delta x/\Delta z$ & 5&.21 & 1&.76 & 1&.95 & 1&.56 & 1&.22 & 0&.953\\\hline
\end{tabular}
\begin{list}{}{}
\item[$^{\mathrm{a}}$] $z_0=\langle z(\tau_{\mathrm{R}}=1)\rangle$
\item[$^{\mathrm{b}}$] $\Delta z$ is the vertical grid resolution
\item[$^{\mathrm{c}}$] $\Delta x$ and $\Delta y$ are the horizontal grid resolution; in all simulations considered here, $\Delta x=\Delta y$ was chosen
\end{list}
\end{table*}
Cool main-sequence stars such as the Sun have thick envelopes in which convective heat transport dominates the other energy transport mechanisms and determines the temperature structure. Although the atmosphere sitting on top of the convective envelope is convectively stable, the motions and the thermal inhomogeneity of the layers below extend into the optically thin layers and affect their 3D structure \citep[see][for a review]{Nordlund09}. The convective motions are correlated with the local temperature and pressure fluctuations, which affect the line opacities and thus the spectral information obtainable from a star. Near-surface convection and the resulting spatial intensity and velocity patterns, which are not directly observable, thus have an impact on stellar spectra. This is in addition to their role in generating and structuring magnetic fields and in chemically mixing the envelope material. The knowledge of the 3D structure is therefore essential if one aims at determining stellar parameters (such as abundances, surface gravity, effective temperature, magnetic field strength and geometry, etc.) from spectra using inversion techniques \citep[see, e.\,g.,][]{RG2,FeH}.\par
Comprehensive three-dimensional (magneto-)hydrodynamic simulations have become an important tool to study solar and stellar convection. For the solar case, where spatially resolved observations are available, the simulations were found to be in excellent agreement with the measurements \citep[e.\,g.][]{StNo98, Nordlund09, TMDP13}. The application of comprehensive 3D convection simulations to stars other than the Sun was pioneered by \citet{ND90a, ND90b, ND90c}. In more recent years, 3D simulations of stellar convection were used for studies of abundances \citep[e.\,g.][]{ABUbsp2,ABUbsp1}, dust formation in very-cool stars \citep[e.\,g.][]{Freytag10}, later evolutionary stages \citep[e.\,g.][]{COB,RG2,RG1}, and the calibration of 1D methods to include 3D effects such as line broadening by micro- and macroturbulence \citep[e.\,g.][]{FeH} or the mixing-length parameter \citep[e.\,g.][]{Lu99,Tramp}. A grid of 37 stellar 3D simulations of main-sequence and giant stars was recently presented by \citet{Tramp13}. First simulations including magnetic fields in stellar near-surface convection were performed by \citet{CS16} and \citet{WDB13}.\par 
In this paper, we consider \texttt{MURaM} simulations of non-magnetic convection in six main-sequence stars for a systematic parameter study of near-surface convection in cool main-sequence stars. The simulation setup was chosen as similar as possible for the six stars, in order to facilitate a comparison between the simulation results. Here, we mainly focus on the general characteristics of the non-magnetic convection and atmospheric structure, such as flow geometries, energy fluxes, and mean profiles of temperature, density, and other quantities. In subsequent papers of this series, we will study the morphology of the granulation patterns and its influence on line shapes and the effects of magnetic fields.\\

%
\section{Simulations}
\subsection{The \texttt{MURaM} Code}\label{sec:code}
The \texttt{MURaM} ({\bf M}PS/{\bf U}niversity of Chicago {\bf Ra}diative {\bf M}HD) code was developed by the MHD simulation groups at the Max Planck Institute for Solar System Research (MPS) in Katlenburg-Lindau and at the University of Chicago \citep{Vogd, Vog05}. It solves the three-dimensional, time-dependent MHD equations for a compressible and partially ionised plasma. The non-grey radiation transport included in the code is based on the short-characteristics scheme \citep{Kunasz} and uses opacity binning \citep{Nor82} based on opacity distribution functions computed with \texttt{ATLAS9} \citep{ATLAS9}. For the simulation of the G2V star, the opacity bins previously obtained for the
solar \texttt{MURaM} simulations \citep[see][]{Vog04} were applied. The
opacity bins for the G2V simulation were used in low horizontal resolution
simulations to obtain the approximate stratification for the K0V and F3V stars.
The $\tau$-sorting method was applied to horizontal averages of these atmospheres to produce
a better estimate for the opacity bins. Analogously, the bins of the K0V star were then used
as a starting point for determining opacity binning for the K5V and M0V stars,
and the opacity bins thus obtained for the M0V star were used as the initial
choice for the M2V star. The procedure for each star could be iterated,
however we found that the changes caused by a second iteration were
minimal. For a detailed description of the radiation transfer in the \texttt{MURaM} code, see \citet{Vog04}.\par
The OPAL equation-of-state \citep{opal2, opal1}  for a plasma with a solar composition \citep{AnGr89} was used to generate look up tables for temperature, pressure, and entropy as a function of density and internal energy density. The \texttt{MURaM} code uses discretised spatial derivatives (4th-order centred differences) on a three-dimensional Cartesian grid and explicit time stepping (4th-order Runge-Kutta solver). In order to stabilise the scheme, artificial diffusivities are introduced. The code version used for the simulations discussed here employed a Minmod slope limiter which sets the numerical diffusivity of the scheme \citep[see][]{MUR3}. 
The bottom boundary was open to vertical inflows and arbitrarily inclined outflows. The pressure at the bottom boundary was adjusted to maintain a fixed mass in the computational domain. For the simulations dicussed here, the entropy density of the material entering the domain through the bottom boundary was varied in an iterative process until the desired value of the effective temperature was (approximately) reached. In all six simulations, the top boundary was closed for flows. Test simulations with open top boundary were also run and yielded no significantly different results. The code uses periodic side boundary conditions.\par
\subsection{Stellar parameters}
\begin{figure}
\centering
\includegraphics[width=8.5cm]{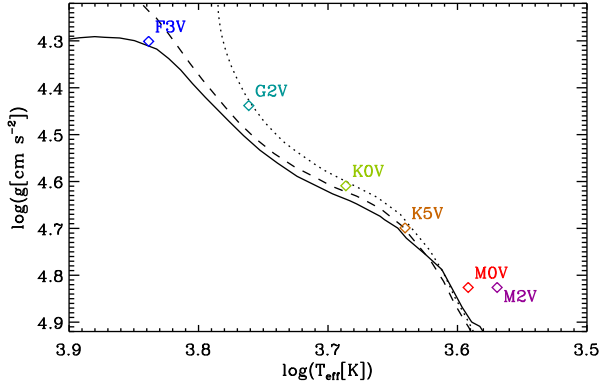}
\caption{Stellar parameters of the six models along with three isochrones by \citet{evotracks}, solid line: zero-age main sequence (ZAMS), dashed line: age of 1 Ga, dotted line: age of 4.5 Ga (approximate solar age) on the $\log g$-$\log T_{\mathrm{eff}}$ plane.}\label{fig:HRD}
\end{figure}
\begin{figure*}
\centering
\begin{tabular}{cc}
\multicolumn{1}{c}{\bf F3V} & \multicolumn{1}{c}{\bf G2V}\\[1mm]
\includegraphics[width=7.8cm, trim=0cm 0cm 0.45cm 0cm]{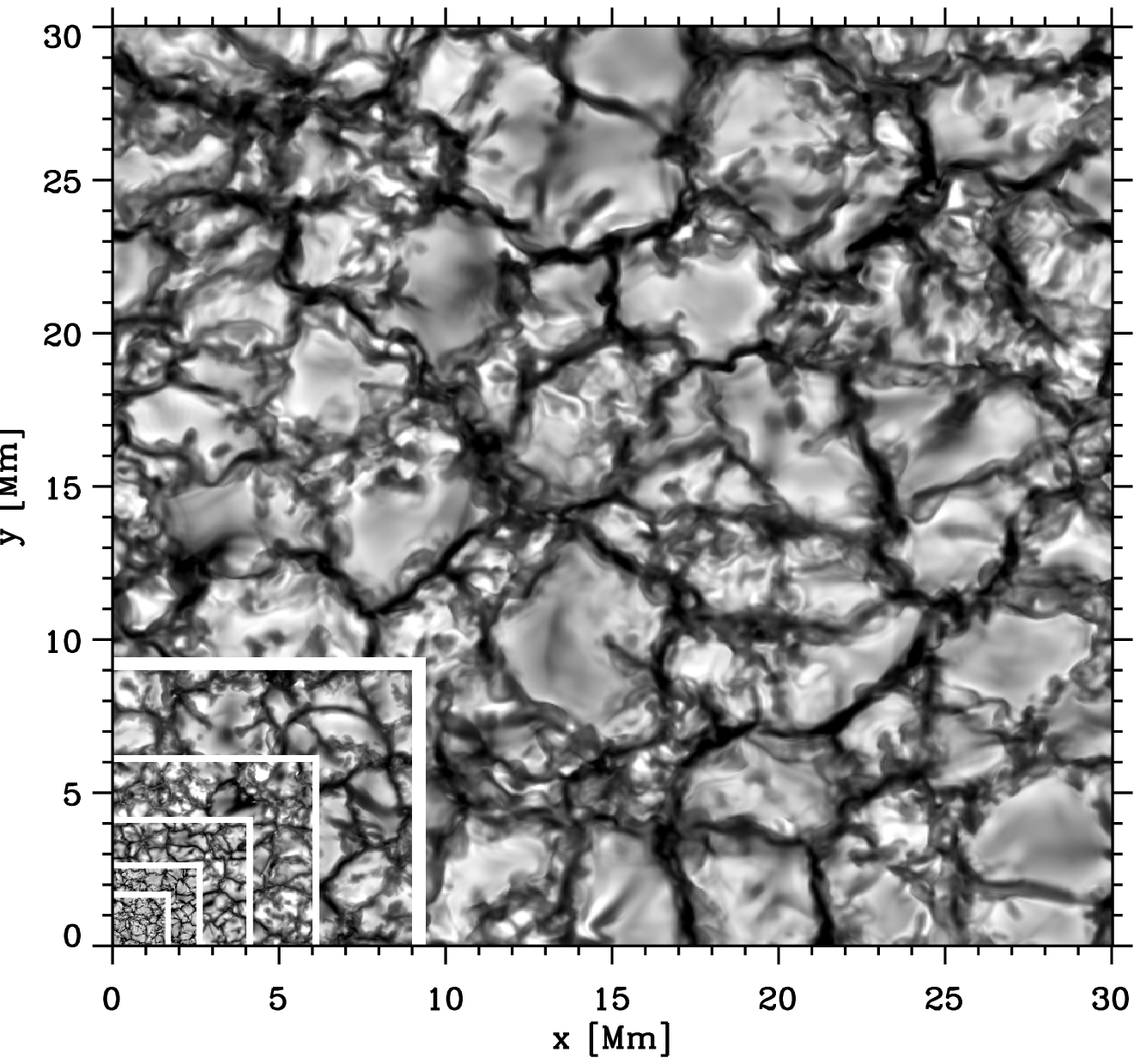} & \includegraphics[width=7.8cm, trim=0cm 0cm 0.45cm 0cm]{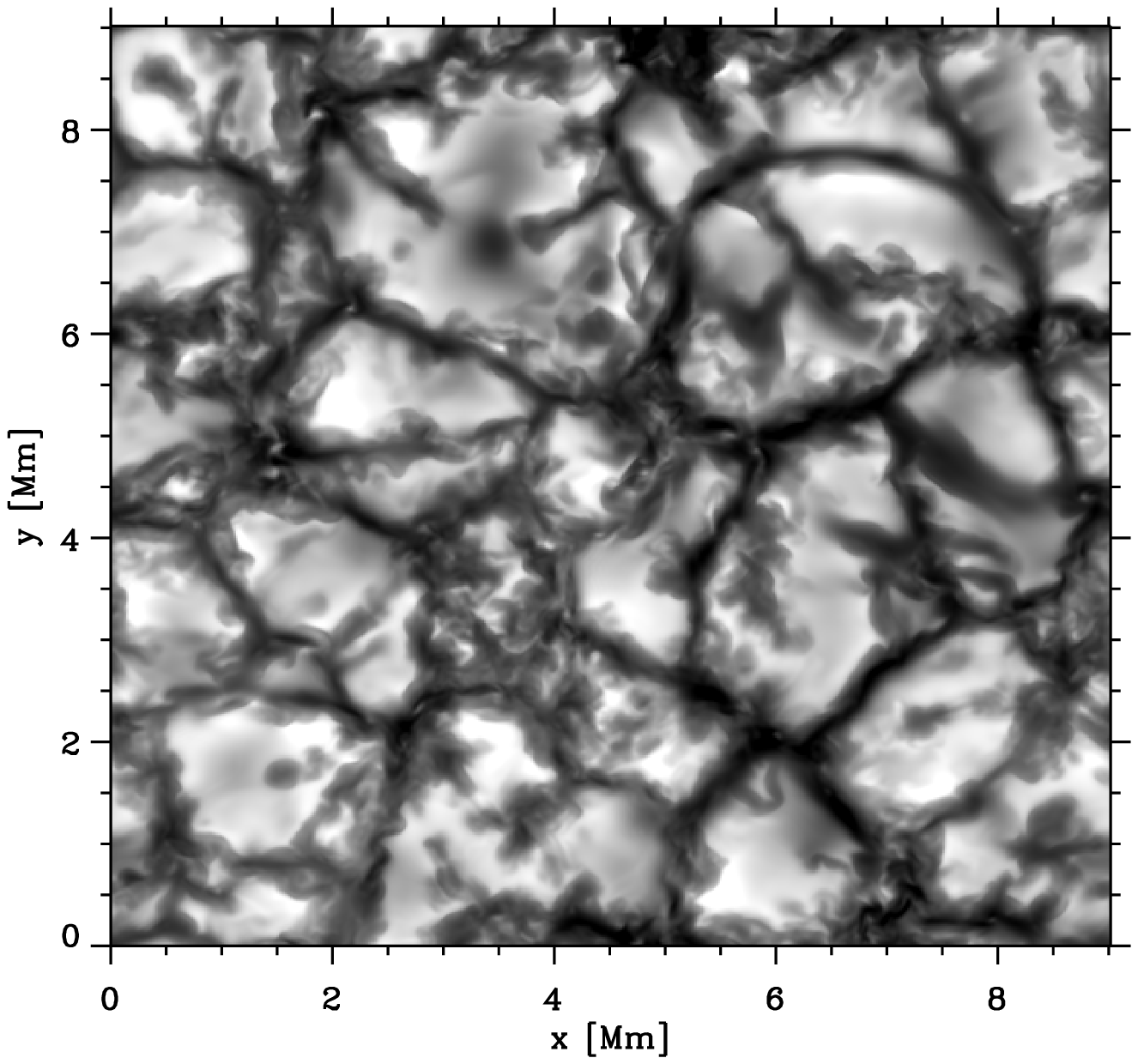}\\[1mm]
\multicolumn{1}{c}{\bf K0V} & \multicolumn{1}{c}{\bf K5V}\\[1mm]
\includegraphics[width=7.8cm, trim=0cm 0cm 0.45cm 0cm]{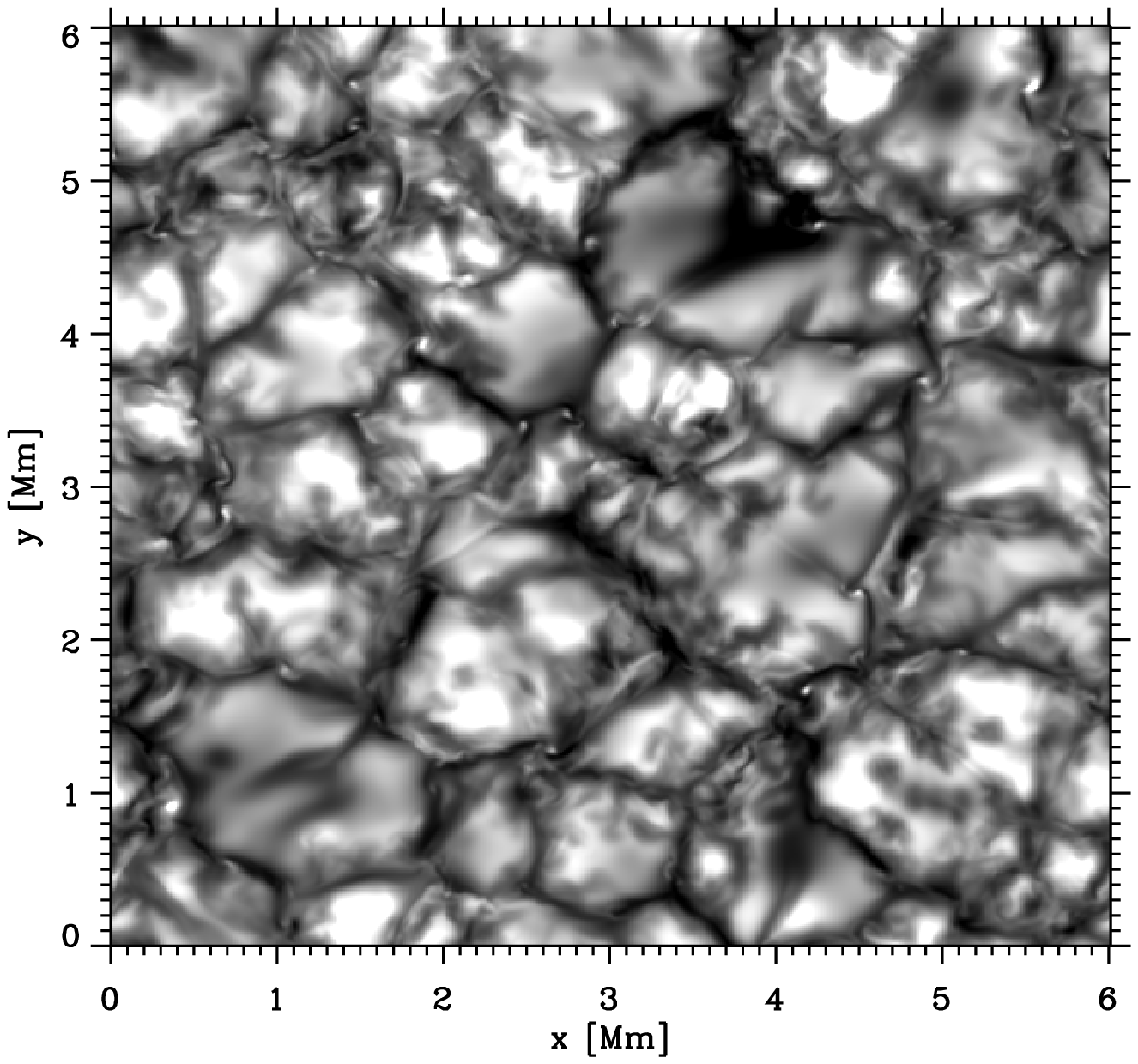} & \includegraphics[width=7.8cm, trim=0cm 0cm 0.45cm 0cm]{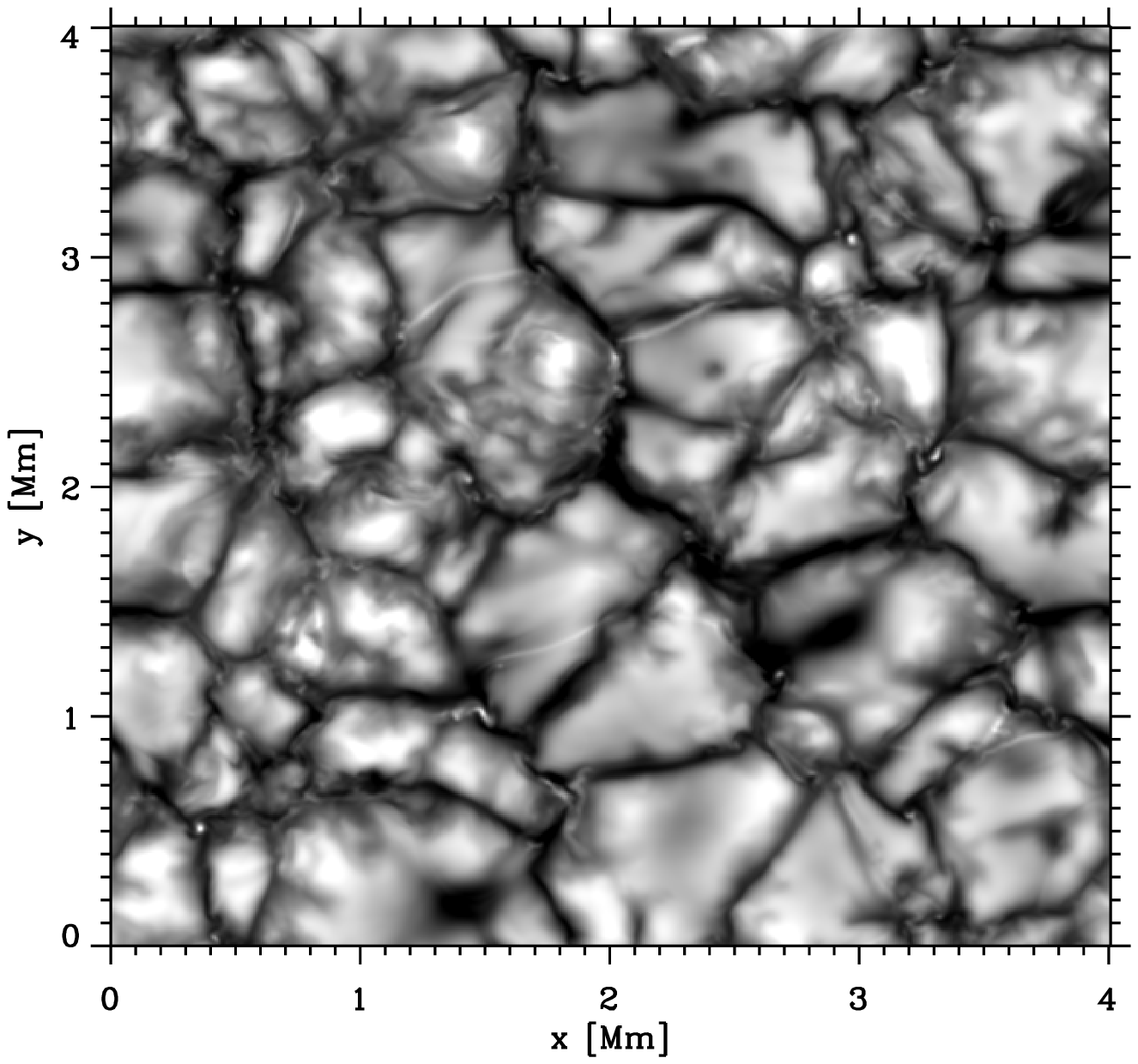}\\[1mm]
\multicolumn{1}{c}{\bf M0V} & \multicolumn{1}{c}{\bf M2V}\\[1mm]
\includegraphics[width=7.8cm, trim=0cm 0cm 0.45cm 0cm]{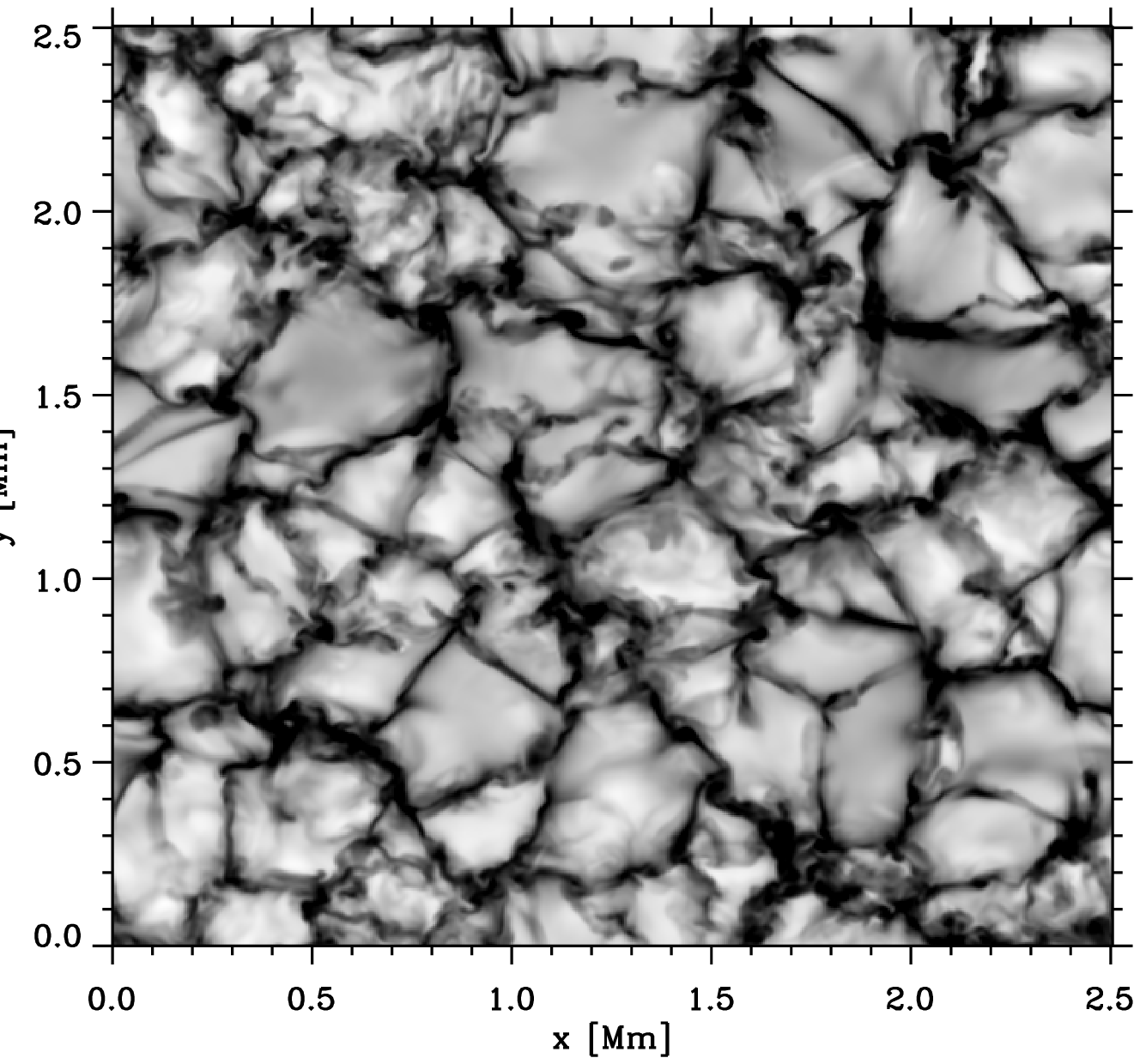} & \includegraphics[width=7.8cm, trim=0cm 0cm 0.45cm 0cm]{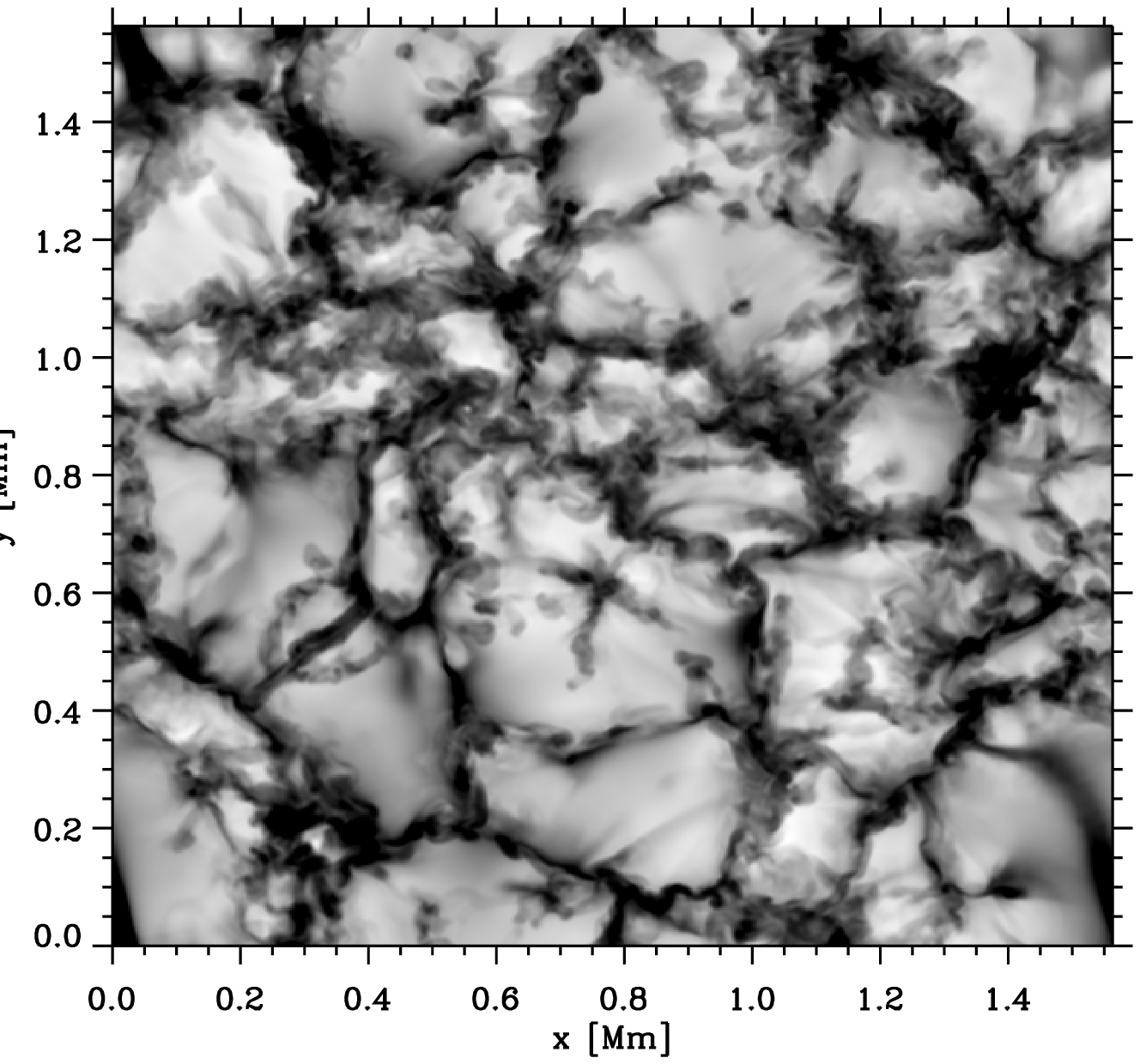}\\[1mm]
\end{tabular}
\caption{Maps of bolometric intensity emerging vertically from single snapshots of the six simulations. The grey scale of each image is saturated at $\pm 2\,\sigma_I$, where $\sigma_I$ is the rms contrast (cf. Table~\ref{tab:values}). The significant difference in the length scales of the images is illustrated by the inset in the {\it upper left panel}, which shows all other images on the same scale as the one from the F3V simulation.}\label{fig:snapshots}
\end{figure*}
For the near-surface layers and atmosphere of a cool star, the governing parameters are the gravitational acceleration, $g$, the effective temperature, $T_{\mathrm{eff}}$, and the chemical composition. We use solar abundances in all cases. The effective temperature and the gravitational acceleration were chosen to match the conditions in cool main-sequence stars. We have carried out simulations corresponding to the following spectral types: F3V, G2V, K0V, K5V, M0V, and M2V (stellar parameters given in table~\ref{tab:values}). Figure\,\ref{fig:HRD} shows the location of the six models in the $\log g$-$\log T_{\mathrm{eff}}$ plane along with three isochrones marking the position of the main sequence \citep{evotracks}.\par
While gravitational acceleration and chemical composition explicitely enter the simulations as parameters, the effective temperature is indirectly specified through the bottom boundary condition of the code, see Sect.~\ref{sec:code}. For the analysis presented here, the simulations have been run long enough with fixed inflowing entropy density for any transients to dissappear, however $T_{\mathrm{eff}}$ varies slightly due to oscillations and granulation. The standard deviation of the temporal fluctuations of $T_{\mathrm{eff}}$ is given in table~\ref{tab:values}.\par
\subsection{Simulation setup}
The dimensions of the computational domain (``local box'') were adapted according to the stellar parameters: the height of the box was chosen to contain about 13 to 15 pressure scale heights (at least six below and six above the optical surface). The vertical cell size (height resolution $\Delta z$) was set sufficiently small to resolve steep temperature gradients and to maintain $\Delta z < H_p/5$ at every depth, where $H_p$ is the local pressure scale height. For most of the models, 300 or less cells in the vertical direction were sufficient to meet these criteria. Only the F3V model with strongly varying pressure scale height and a very steep local temperature gradient, required 800 cells in the vertical direction. The two equal horizontal box dimensions were scaled to the expected granule (convection cell) size. In order to reduce the effects of the periodic boundary conditions and obtain good statistics while maintaining sufficient spatial grid resolution, the boxes were chosen big enough to contain 30 to 50 granules at any given time. The horizontal dimensions were resolved into $512\times 512$ grid cells. The largest computational domain of the simulations thus comprised $512 \times 512 \times 800\approx 2.1\cdot 10^8$ cells. Table~\ref{tab:H_p} gives a summary of the computational box dimensions and grid resolutions.\par

%
\section{Results}
\subsection{General morphology of near-surface convection}
Figure~\ref{fig:snapshots} gives maps of the bolometric intensity emerging vertically from the simulated stellar surfaces for single snapshots of the time-dependent simulations. All simulations show intensity patterns reminiscent of solar granulation. The typical size of the granules varies from $\sim$5\,Mm for F3V to $\sim$0.3\,Mm for M2V. The rms bolometric intensity contrast (denoted by $\sigma_I$ in Table\,\ref{tab:values}) decreases from about 20\% for F3V to less than 3\% in the M2V simulation, reflecting decreasing temperature fluctuations on surfaces of constant optical depth (see Sect.~\ref{sec:strat}).\par
There are qualitative changes in the visual appearance of the surface convection along the sequence of simulated stars. For instance, the granulation pattern of the F3V model appears ``rough'' and irregular owing to numerous shock waves at the optical surface. Shocks are rarer and weaker in the near-surface layers of the cooler stars since the typical convective velocities are lower (also in relation to the sound speed; cf. Fig.~\ref{fig:v}). At the cool end of our model sequence, the M-dwarf granules, which are sustained by the slowest convective flows, have more irregular shapes but less brightness substructure than their counterparts on the simulated G- and K-type stars. As we report quantitatively in Paper II, their dark intergranular lanes are thinner (with respect to the granule size) and vary more strongly in intensity and width than those of the other stars \citep[see also][]{Lu02}.\par
\begin{figure}
\centering
\begin{tabular}{cc}
{\bf F3V} & {\bf G2V}\\
\includegraphics[width=0.45\columnwidth]{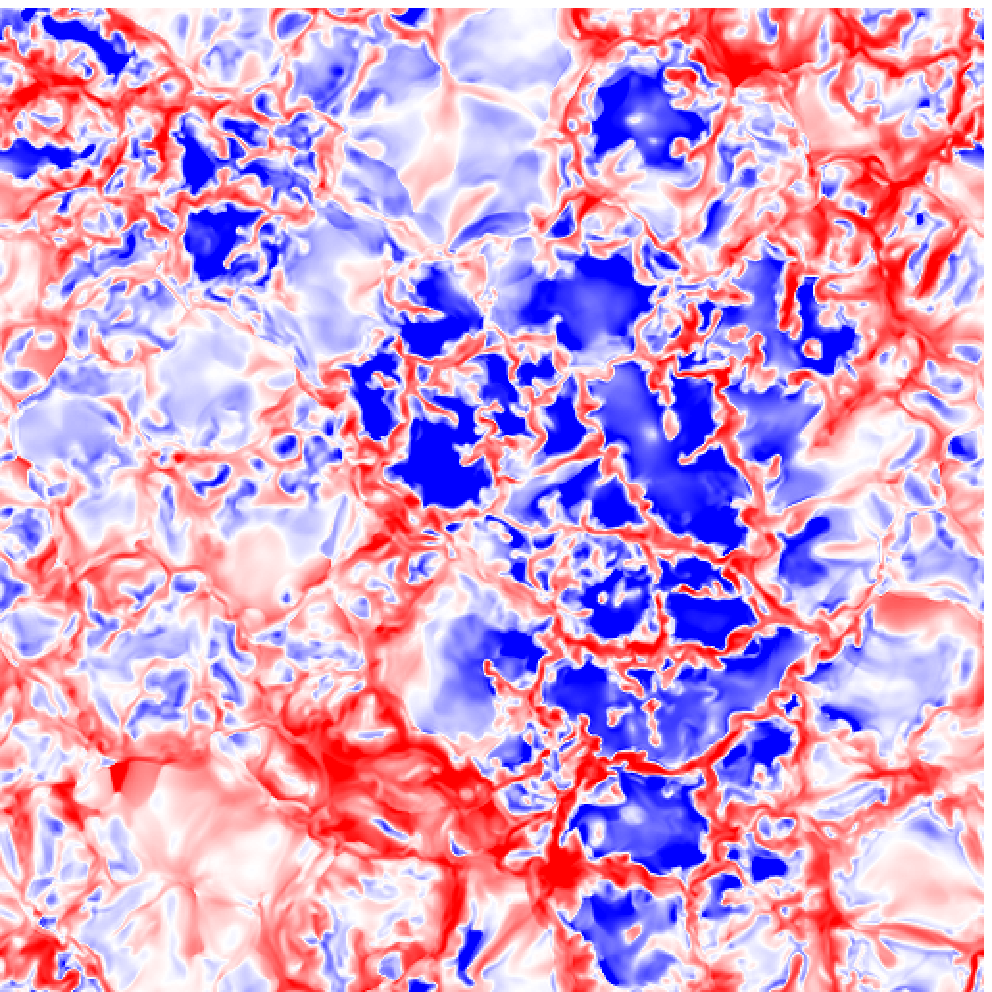} & \includegraphics[width=0.45\columnwidth]{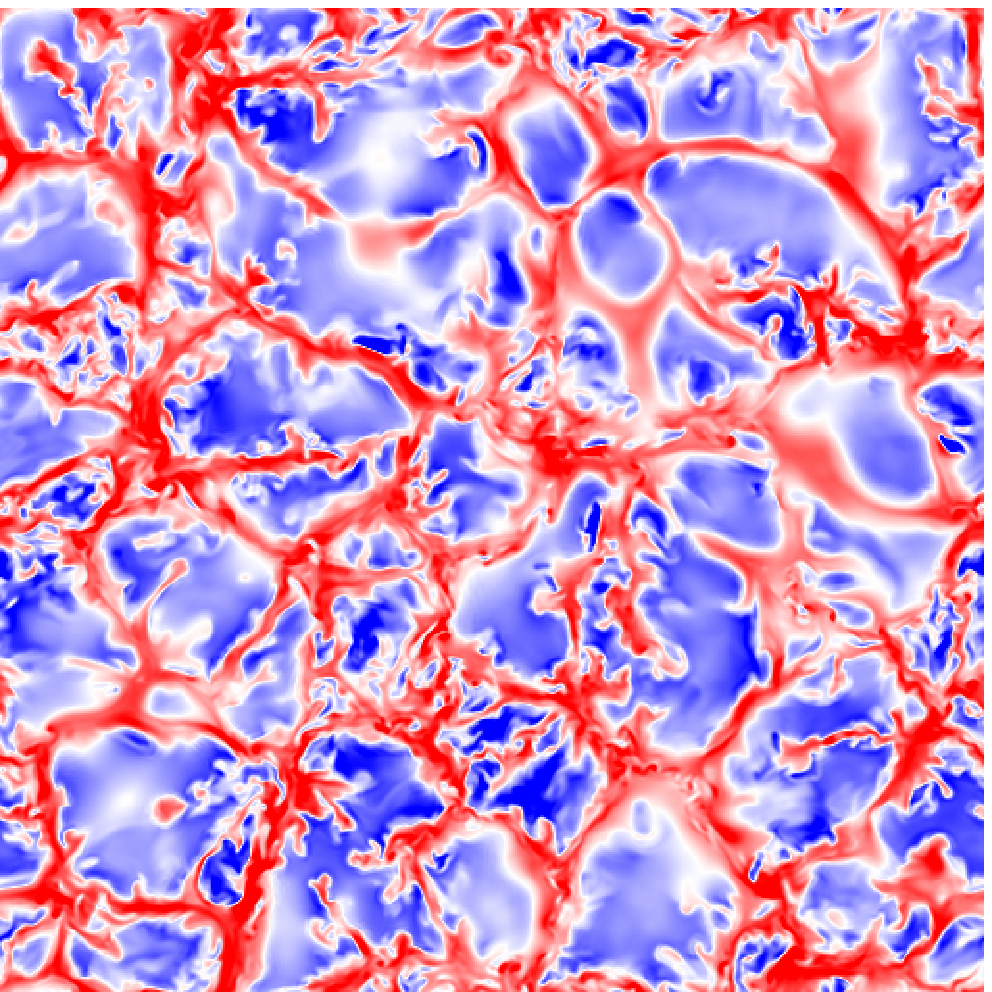}\\
{\bf K0V} & {\bf M0V}\\
\includegraphics[width=0.45\columnwidth]{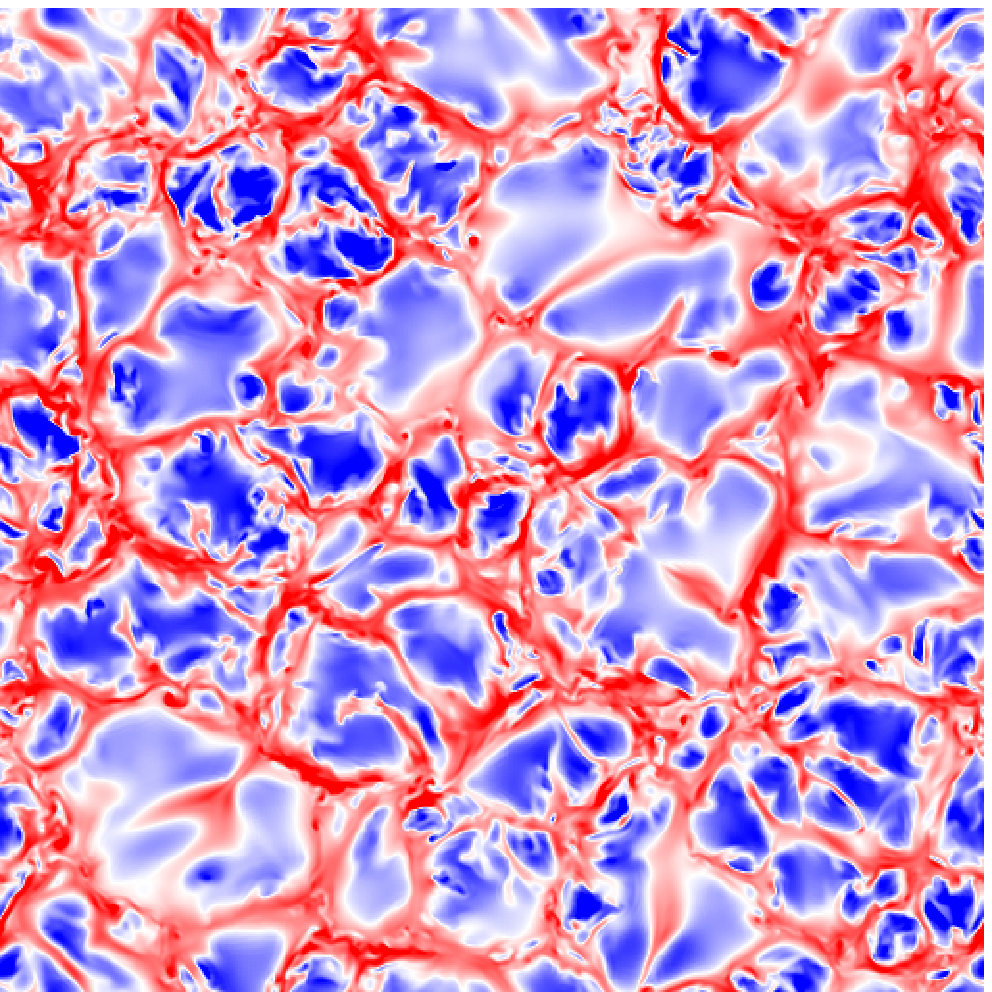} & \includegraphics[width=0.45\columnwidth]{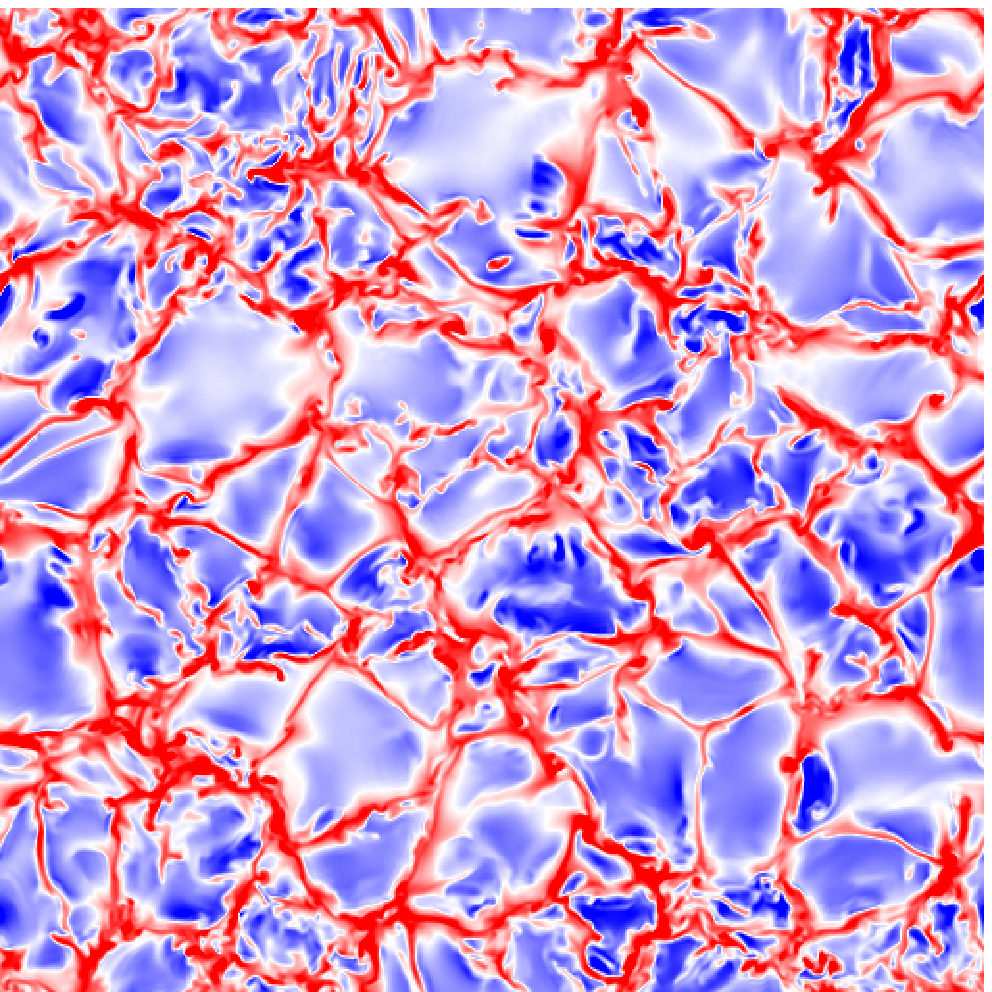} \\
\end{tabular}
\caption{Maps of the vertical velocity $\vel_z$ at constant geometrical depth $z_0=\langle z(\tau_{\mathrm{R}}=1)\rangle$ (average level of the optical surface) for four of the six models. Upward motions are blue, downward motions are red, colour scales saturate at $\pm 2 \cdot \vel_{z,\mathrm{rms}}(z_0)$ (for values, see Table~\ref{tab:values}). Note that the horizontal scales are different (cf. Fig.~\ref{fig:snapshots}).}\label{fig:vz_img0}
\end{figure}
\begin{figure}
\centering
\begin{tabular}{cc}
{\bf F3V} & {\bf G2V}\\
\includegraphics[width=0.45\columnwidth]{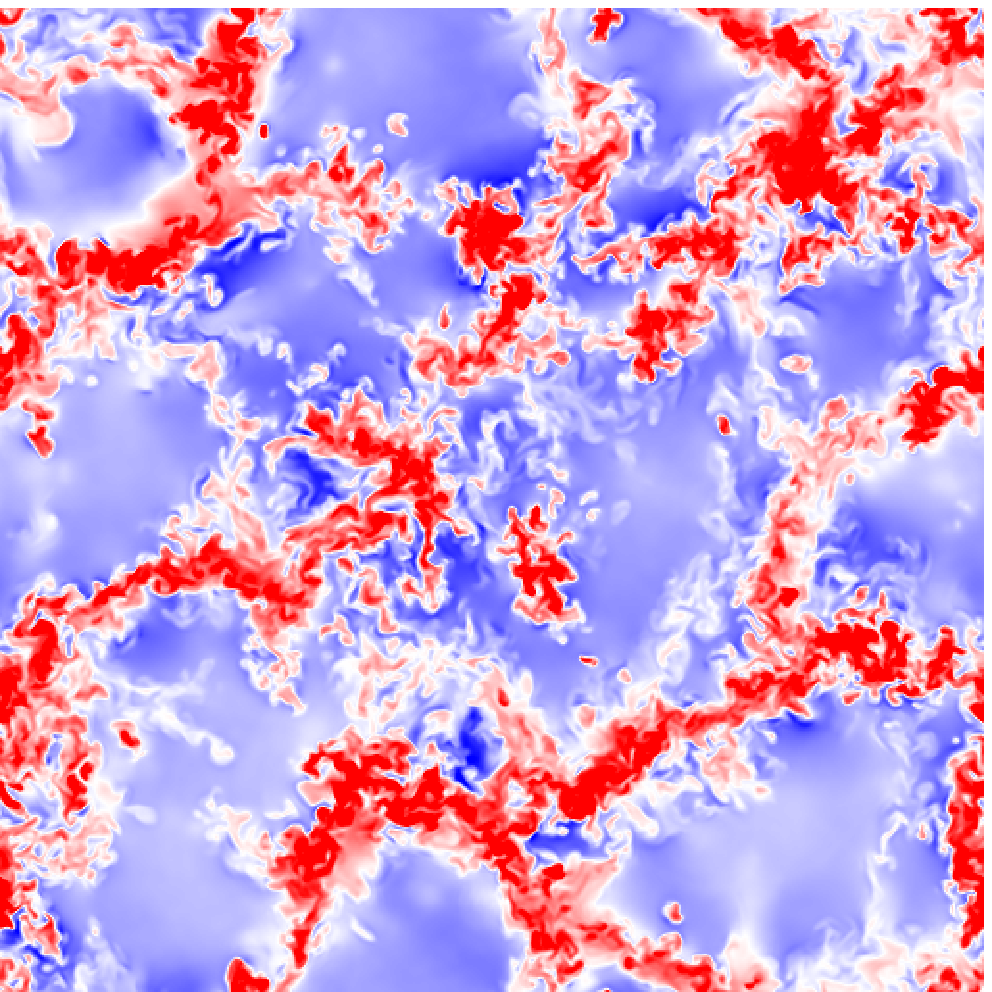} & \includegraphics[width=0.45\columnwidth]{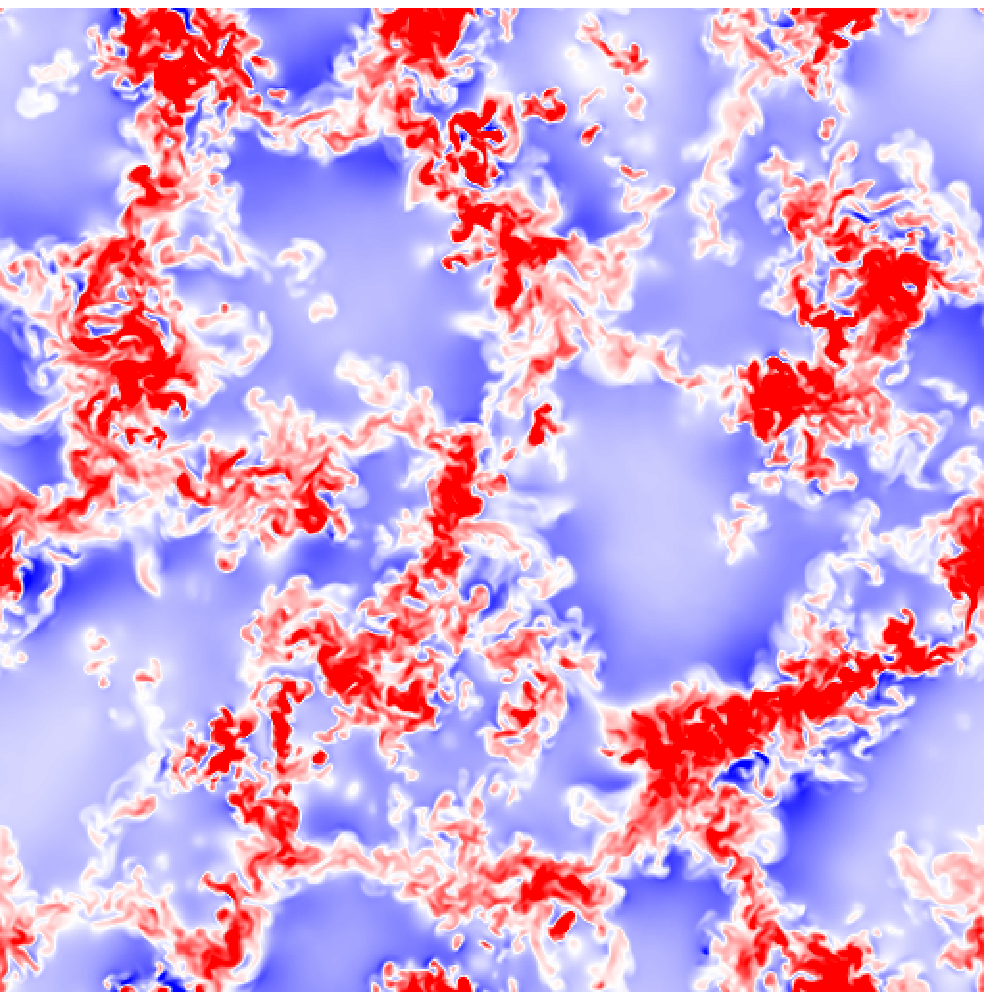}\\
{\bf K0V} & {\bf M0V}\\
\includegraphics[width=0.45\columnwidth]{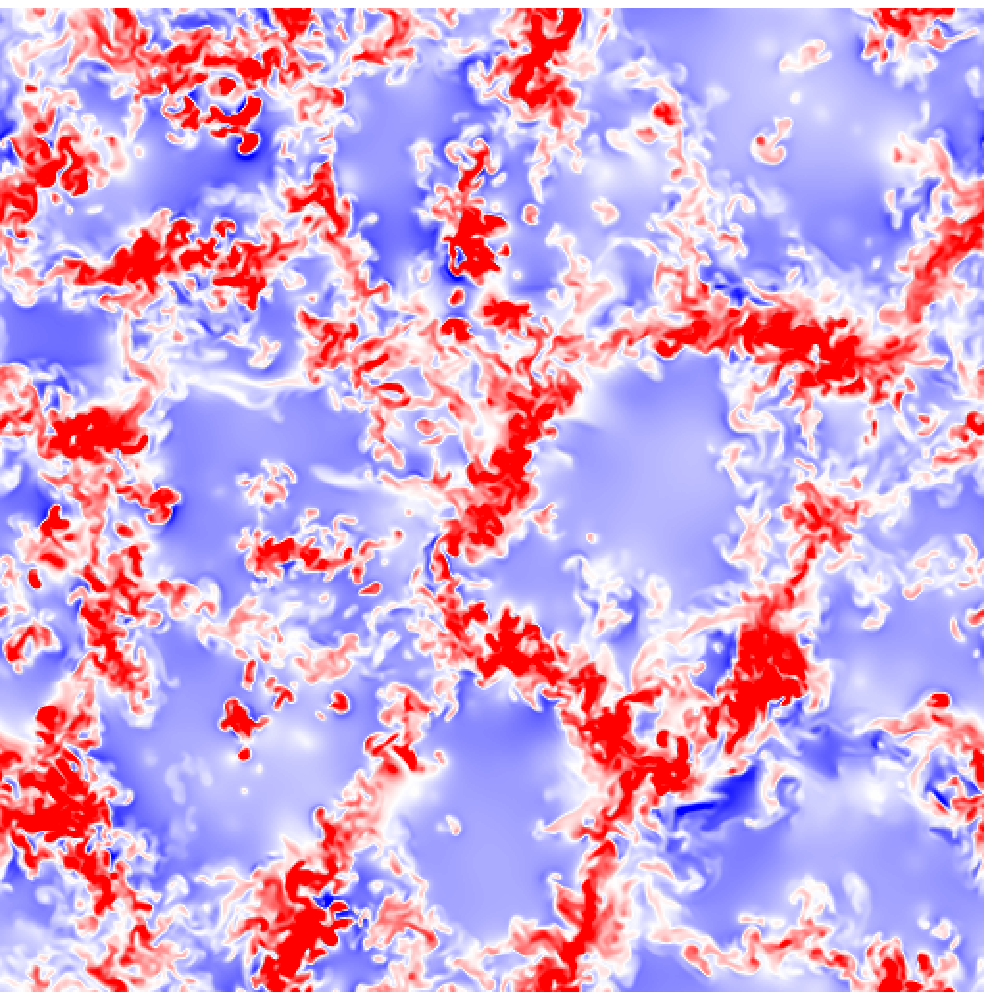} & \includegraphics[width=0.45\columnwidth]{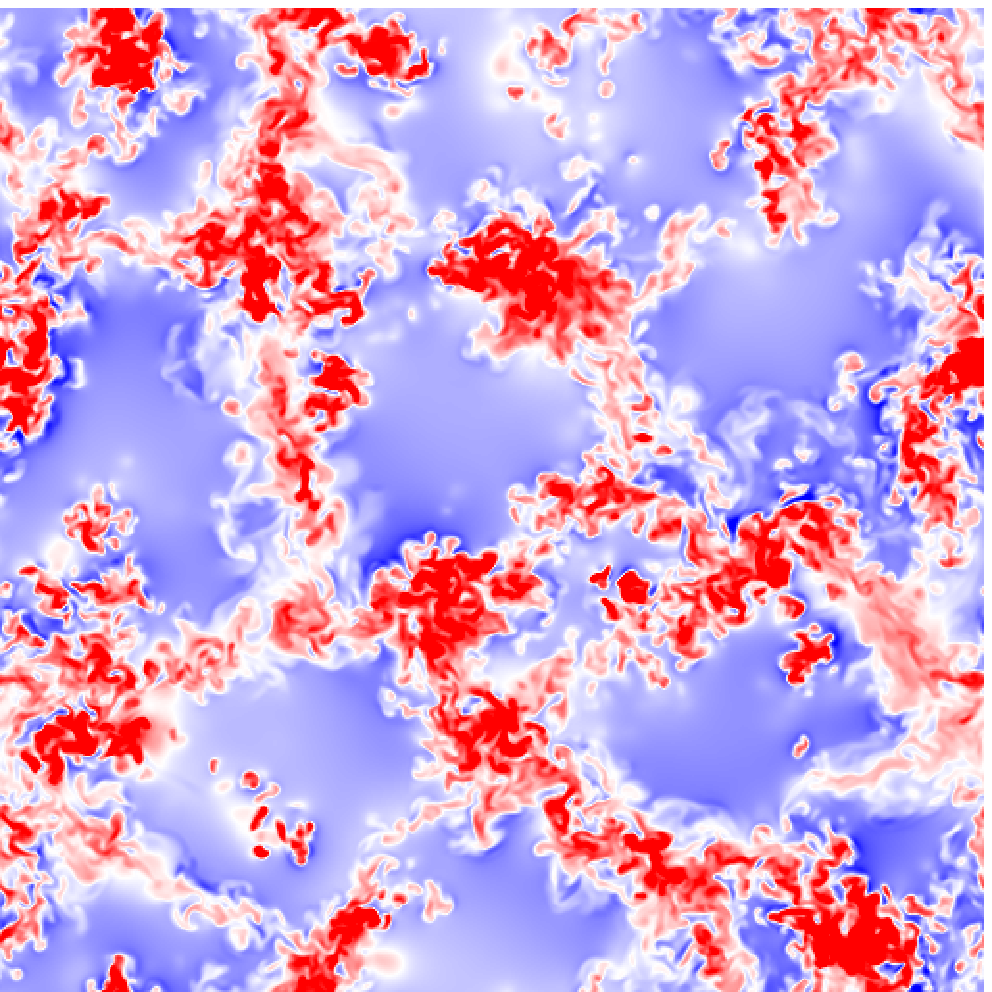}\\
\end{tabular}
\caption{Same as Figure\,\ref{fig:vz_img0}, but for geometrical depth $z_2$ with \mbox{$\langle p(z_2)\rangle=100\,\langle p(z_0)\rangle$}, corresponding to 4.6 pressure scale heights below the optical surface.}\label{fig:vz_img2}
\end{figure}
\citet{Lu06} found ``dark knots'' associated with strong downflows and vortex motion in simulations of convection in M-type main- and pre-main-sequence objects. Our simulations show knots of high vorticity associated with strong downflows in all models (some examples in Fig.\,\ref{fig:snapshots} are: G2V, $(x,y)=(8.7\,\mathrm{Mm}, 4.4\,\mathrm{Mm})$; K5V, $(x,y)=(0.36\,\mathrm{Mm},0.52\,\mathrm{Mm})$; M0V, $(x,y)=(0.45\,\mathrm{Mm},0.6\,\mathrm{Mm})$). They become increasingly stable and prominent at lower effective temperatures. In our models, some of these vortices are evacuated strongly enough by the effect of the centrifugal force to become brighter than their surroundings \citep[cf. vortices in solar simulations studied by][]{Moll11, Moll12}. Most frequently these bright vortex structures occur in our two K-type simulations.\\
A more detailed analysis of the granulation properties and their effects on spectral lines is given in \citet[][Paper II, hereafter]{Paper2}.

\subsection{Velocity field}\label{sec:flows}
As the visible granulation pattern is created by convective flows, it is strongly correlated to the vertical velocities at the optical surface, \mbox{$\mathbf{\vel_z(z=0)}$}. Figure~\ref{fig:vz_img0} shows \mbox{$\mathbf{\vel_z(z=0)}$} for four of the six simulations. The snapshots are taken at the same time as in Figure~\ref{fig:snapshots}. The colour scale of the images saturate at \mbox{$2\,\vel_{z,\mathrm{rms}}(z_0)$} with \mbox{$z_0:=\langle z\rangle_{\tau_{\mathrm{R}}=1}$}, values of which are given in Table~\ref{tab:values}. The granules visible in Figure~\ref{fig:snapshots} correspond to upflows, while the dark intergranular lanes correspond to downdrafts. In the \mbox{G-,} \mbox{K-,} and M-type simulations, an anti-correlation between size and mean upflow velocity of the granules is indicated: while most of the small convection cells appear (almost) saturated in Figure~\ref{fig:vz_img0}, meaning their velocity reaches $2\,\vel_{z,\mathrm{rms}}(z_0)$, the larger granules appear paler, meaning their upflow speed is lower. In the F3V simulation, this effect is not visible, due to a strong large-scale modulation of the vertical velocity at the optical surface. This large scale pattern might hint to a strong mesogranulation in this spectral type. Unfortunately, the length scale of this modulation is the horizontal box size, which raises the question whether this effect is produced, enhanced, or modified by the periodic boundary condition. A test simulation with a box twice as large has shown similar but weaker large-scale modulation but probably is still strongly influenced by the periodic boundary condition.\\
Figure \ref{fig:vz_img2} shows maps of the vertical velocity at a depth of 4.6 pressure scale heights below the optical surface, where the average pressure is 100 times the average pressure at the optical surface, \mbox{$p_0:=\langle p\rangle_{\tau_{\mathrm{R}=1}}$}. The typical size of the convection cells is significantly larger at this depth than at the surface. A rough estimate based on mass conservation and stationarity predicts a proportionality between the horizontal scale of the vertical velocity pattern at a given depth, $D_{\mathrm{hor}}$, and the local density scale height, $H_{\varrho}$ \citep{Nordlund09},
\begin{equation}\label{eqn:Dhor}
D_{\mathrm{hor}}=4\,H_{\varrho}\left(\frac{\vel_{\mathrm{hor}}}{\vel_{\mathrm{ver}}}\right)\,\,,
\end{equation}
where $\vel_{\mathrm{hor}}$ and $\vel_{\mathrm{ver}}$ are the horizontal and vertical convection velocities, respectively. They can be approximated by the horizontally averaged (height-dependent) rms values of the vertical and horizontal components of the fluid velocity,\[ 
\vel_{\mathrm{ver}}\approx\vel_{z,\mathrm{rms}}:=\sqrt{\langle \vel_z^2\rangle_{z}}\]
and
\[\vel_{\mathrm{hor}}\approx\vel_{x,y,\mathrm{rms}}:=\sqrt{\langle \vel_x^2+\vel_y^2 \rangle_z}\,\mathrm{,}\] 
respectively (for the definition and discussion of the horizontal average $\langle\cdot\rangle_z$, see Appendix~\ref{app:avg}). 
In the left panel of Figure~\ref{fig:Dhor} the ratio of $\vel_{x,y,\mathrm{rms}}$ and $\vel_{z,\mathrm{rms}}$ is plotted as a function of normalised gas pressure $\langle p\rangle_z/p_0$. In subsurface layers, we find $\vel_{\mathrm{hor}}/\vel_{\mathrm{ver}}\approx 1$, hence Eq.~(\ref{eqn:Dhor}) predicts that the horizontal scale of the flow pattern roughly follows the trend of the inwardly increasing density scale height.\par
The right panel of Figure\,\ref{fig:Dhor} shows the profile of $D_{\mathrm{hor}}$ as derived from Eq.\,(\ref{eqn:Dhor}) in units of the horizontal box size $X_{\mathrm{tot}}$ of the respective simulation. The density scale height obtained from the simulations was smoothed (convolution with a Gaussian kernel $\sigma=10\Delta z$) to avoid a sharp maximum of $D_{\mathbf{hor}}$ at the optical surface of the two hottest models (cf. Fig.~\ref{fig:scaleheights}). The predicted horizontal scale of 12 -- 20\,\% of the horizontal box size at the optical surface and 25 -- 40\,\% of the horizontal box size at $p=100\cdot p_0$ matches the sizes of the patterns visible in Figures~\ref{fig:vz_img0} and \ref{fig:vz_img2}.\par
\begin{figure*}
\centering
\includegraphics[width=8.5cm]{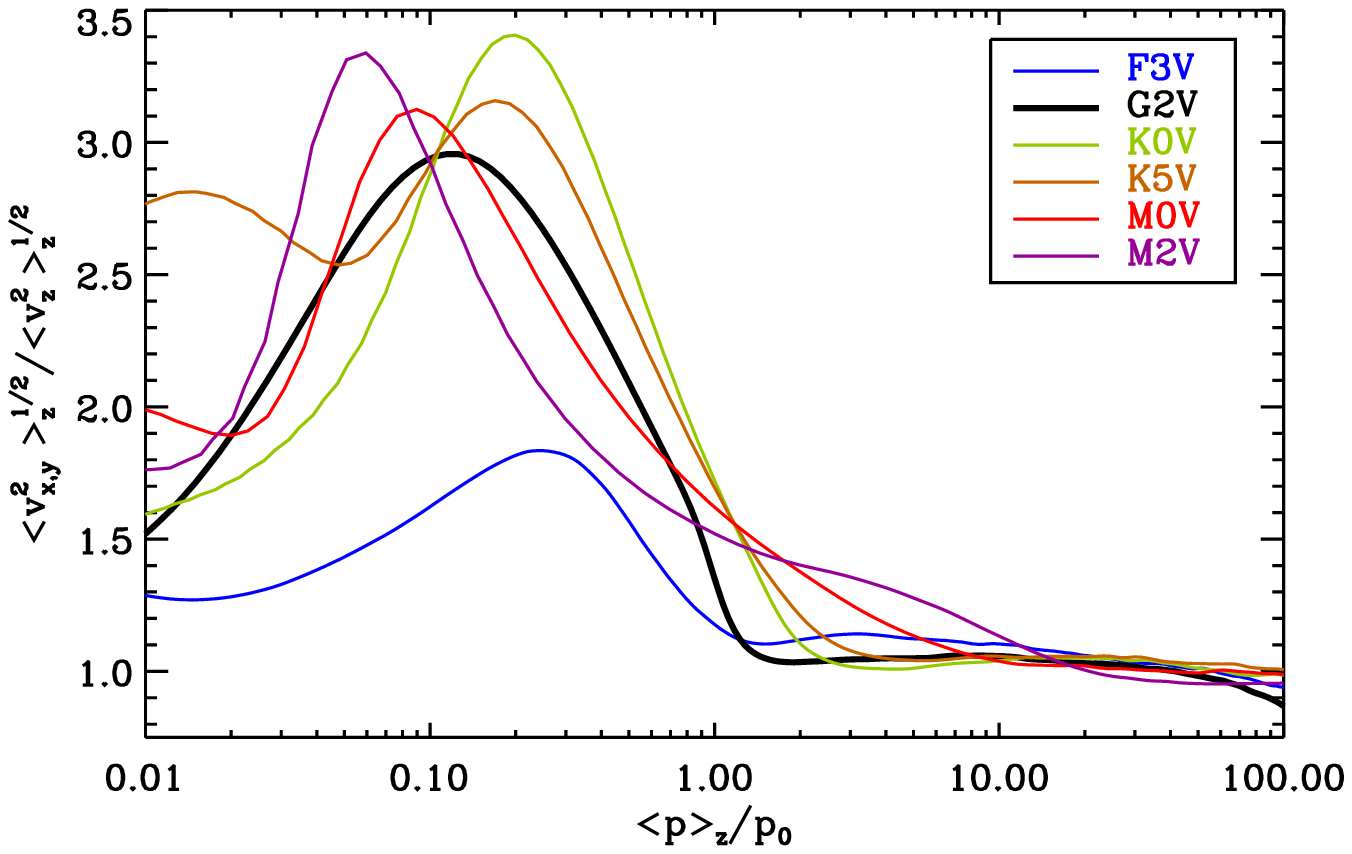}~%
\includegraphics[width=8.5cm]{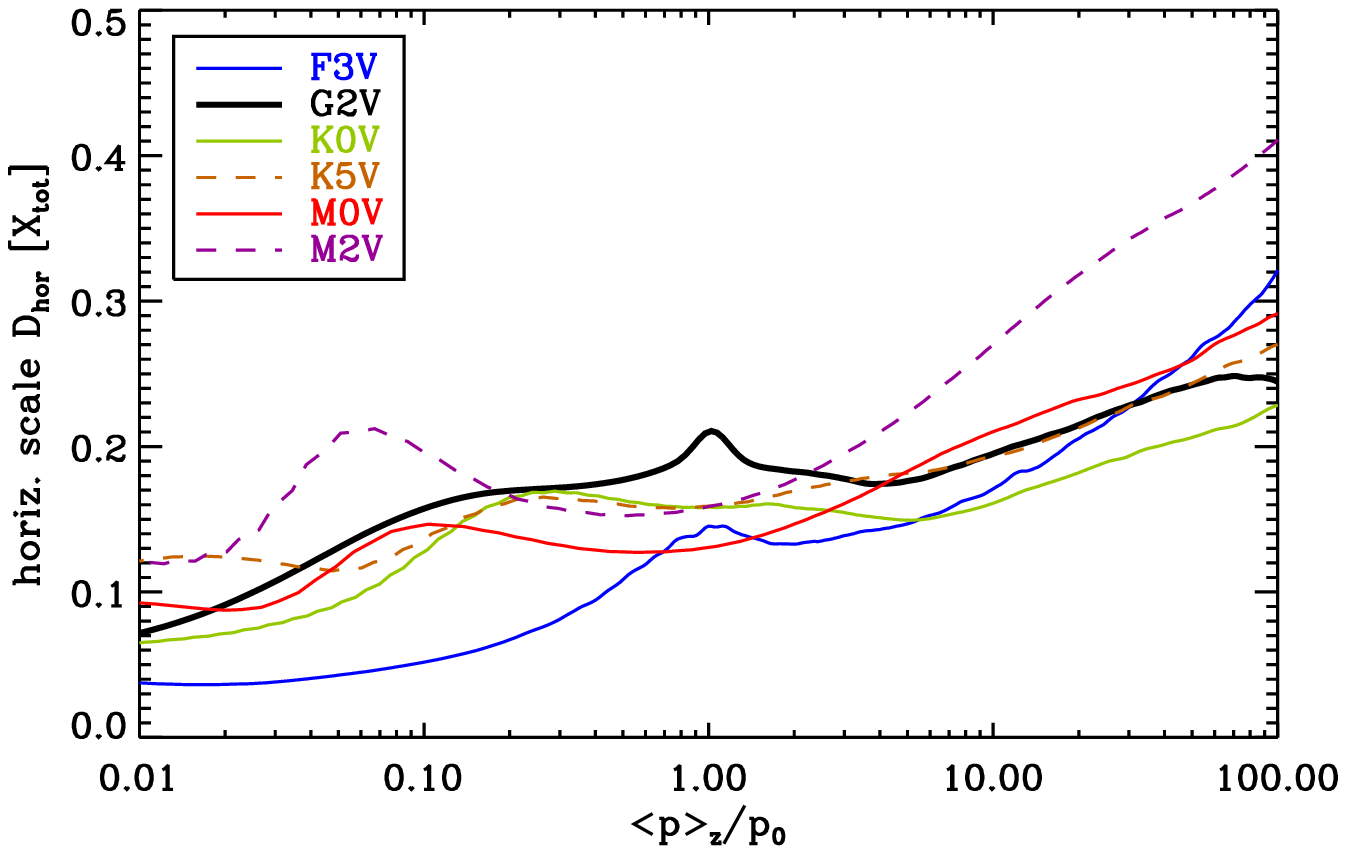}
\caption{Properties of the convective flows. {\it Left}: ratio of rms values of horizontal and vertical flow velocity on surfaces of constant geometrical depth. {\it Right}: estimated horizontal scale of the vertical velocity patterns as derived from Eq.~(\ref{eqn:Dhor}) on surfaces of constant geometrical depth. The horizontal scale is given in units of the horizontal box size, $X_{\mathrm{tot}}$, for an easier comparison with Figs.~\ref{fig:vz_img0} and \ref{fig:vz_img2}. The solid curves refer to the four simulations shown in these figures, the dashed curves to the remaining two simulations.}\label{fig:Dhor}
\end{figure*}
\begin{figure*}
\centering
\includegraphics[width=8.5cm]{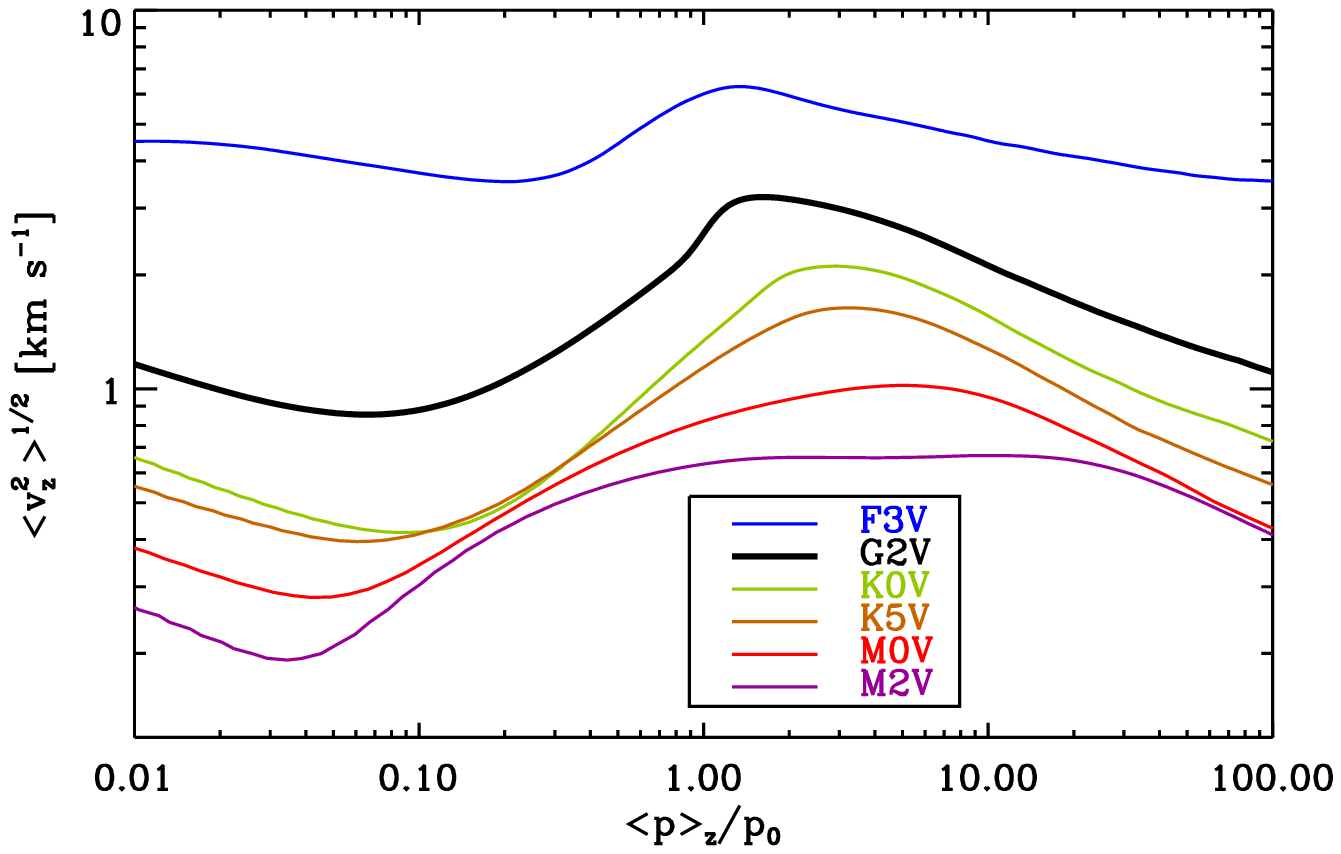}~%
\includegraphics[width=8.5cm]{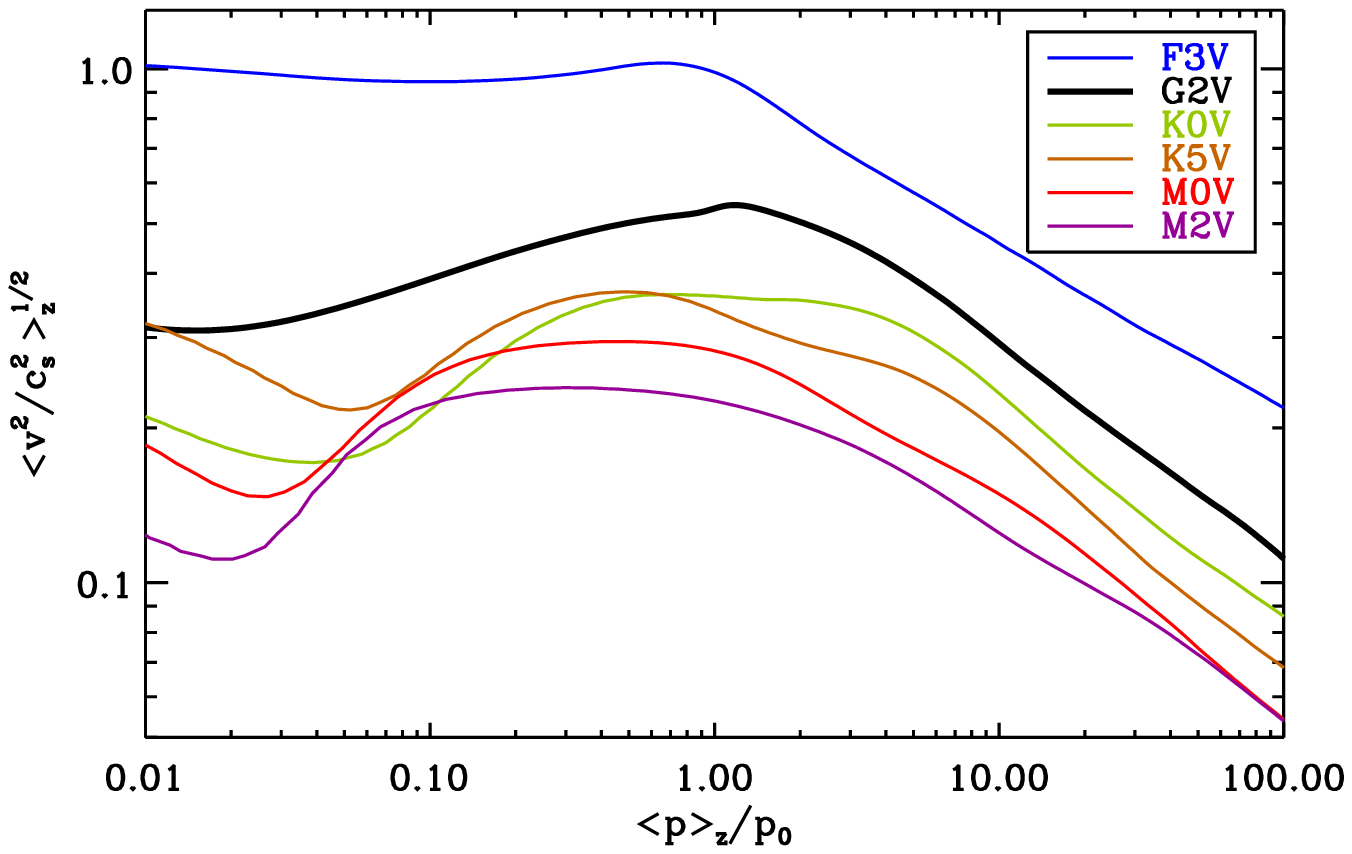}
\caption{Flow velocity rms.~{\it Left}: rms of the vertical component of the flow velocity on surfaces of constant geometrical depth. {\it Right}: rms of the modulus of the flow velocity in units of the local sound speed, $c_s$, (Mach number) on surfaces of constant geometrical depth.}\label{fig:v}
\end{figure*}
\begin{figure*}
\centering
\includegraphics[width=8.5cm]{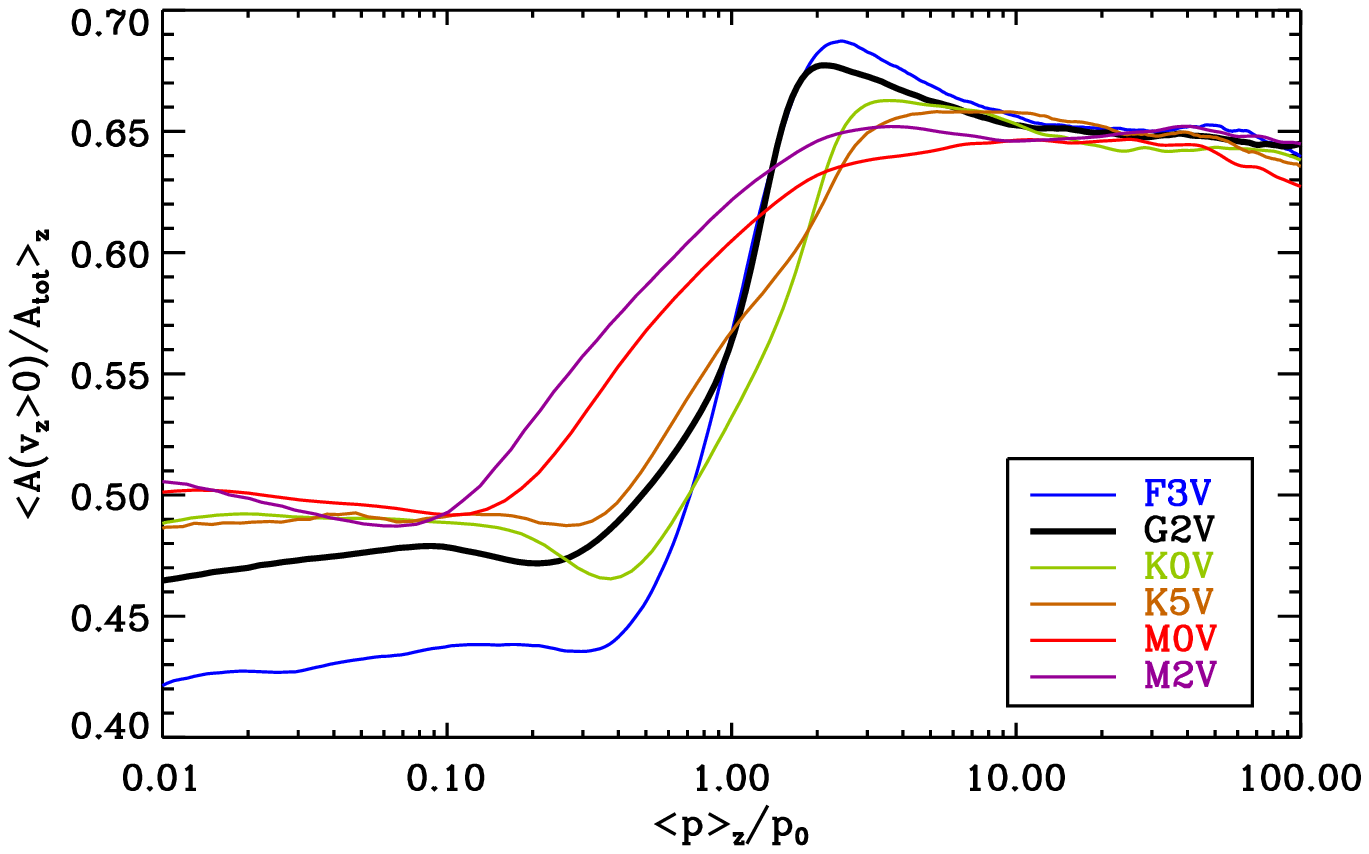} \includegraphics[width=8.5cm]{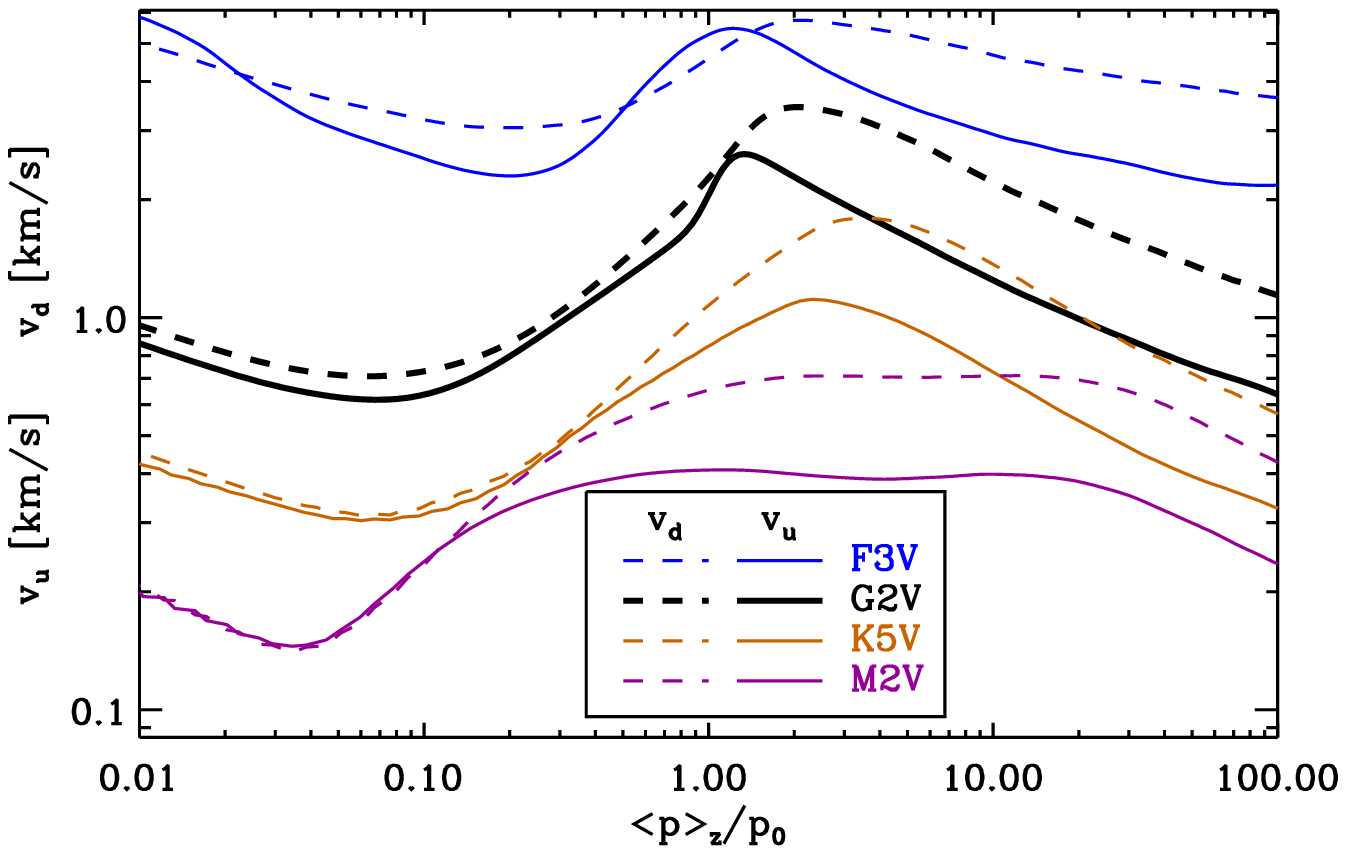}
\caption{Properties of up- and downflows.~{\it Left}: relative area covered by upflows ($\vel_z>0$) on surfaces of constant geometrical depth as functions of normalised averaged pressure. {\it Right}: average speed of the upflows, $\vel_u:=\langle \vel_z|_{\vel_z>0}\rangle_{z}$, (solid) and of the downflows, $\vel_d:=\langle \vel_z|_{\vel_z<0}\rangle_z$, (dashed) as functions of normalised average pressure for four of the six simulations.}\label{fig:upanddown}
\end{figure*}
\comment{\begin{figure}
\centering
\includegraphics[width=8.5cm]{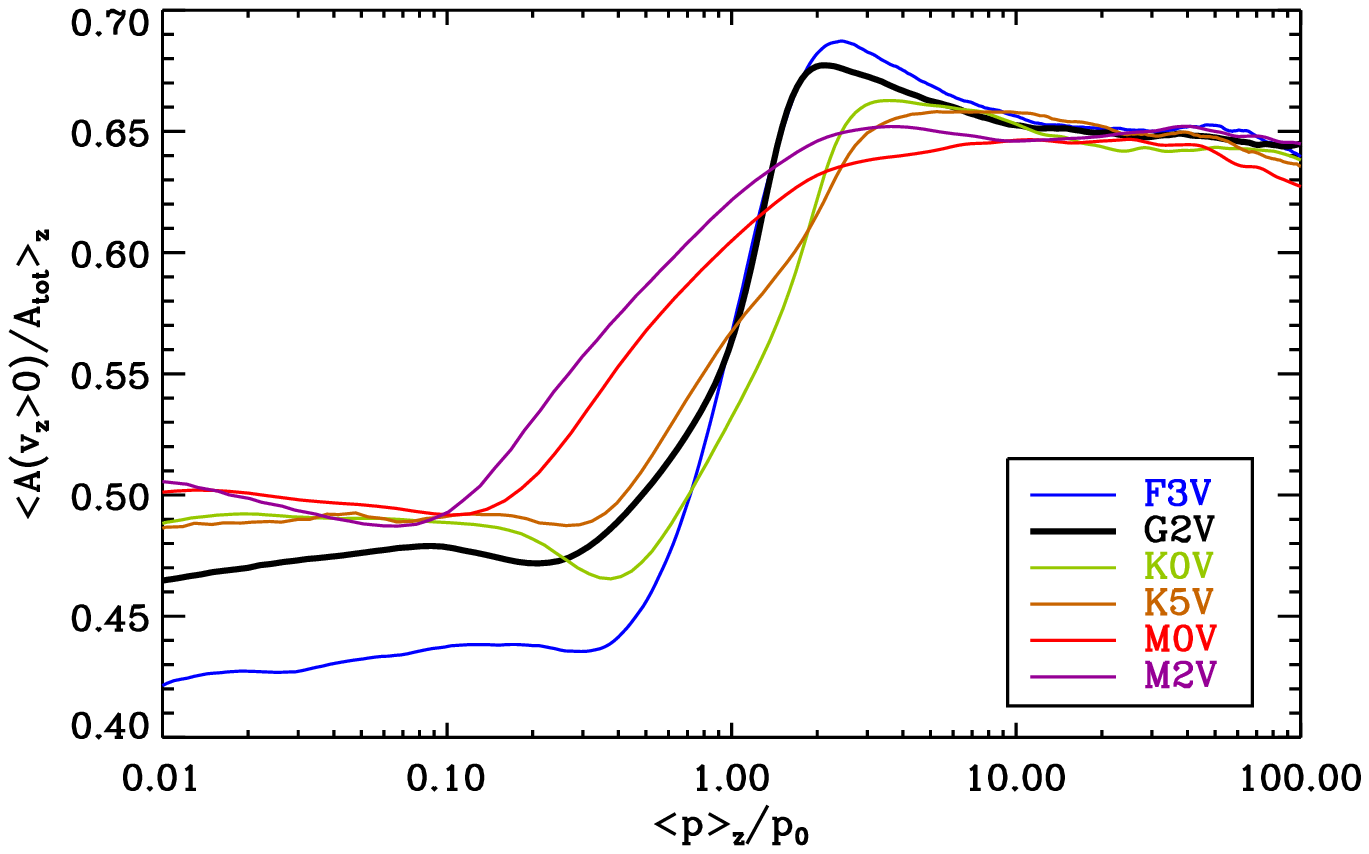}\\
\includegraphics[width=8.5cm]{VU_strat_nP.eps}\\
\includegraphics[width=8.5cm]{VD_strat_nP.eps}
\caption{{\it Top}: relative area covered by upflows ($\vel_z>0$) on surfaces of constant geometrical depth as functions of normalised averaged pressure. {\it Middle}: average speed of the upflows (on surfaces of constant geometrical depth) as functions of the normalised pressure. {\it Bottom}: average speed of the downflows (on surfaces of constant geometrical depth).}\label{fig:upanddown}
\end{figure}}

\begin{figure*}
\centering
\includegraphics[width=8.5cm]{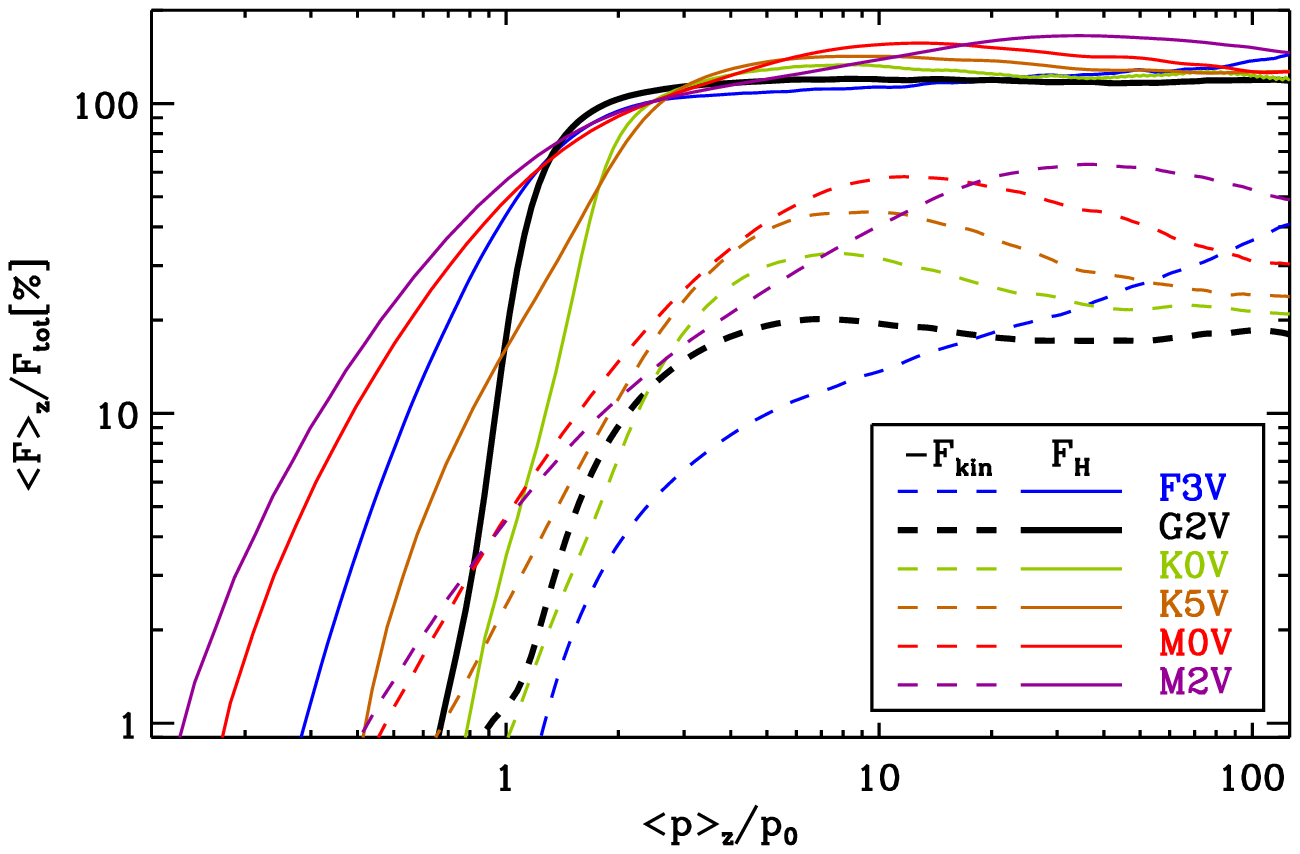}~%
\includegraphics[width=8.5cm]{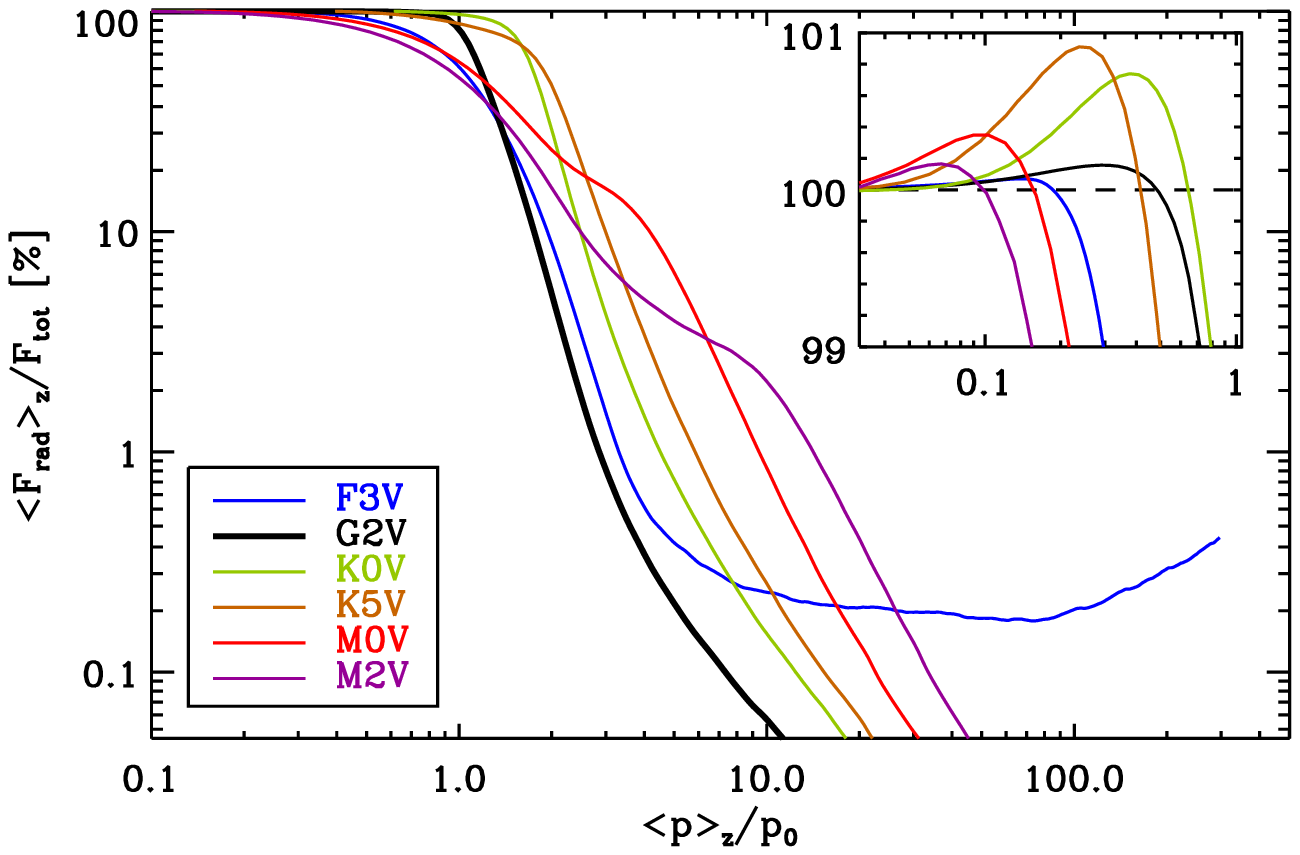}
\caption{Profiles of average energy flux, normalised to total flux. {\it Left:} convective energy flux split into net enthalpy flux and net kinetic energy flux. {\it Right:} radiative flux.}\label{fig:strat6}
\end{figure*}

The left panel of Figure~\ref{fig:v} shows the depth dependence of $\vel_{z,\mathrm{rms}}$, which is a measure of the typical convective velocity. The profiles of $\vel_{z,\mathrm{rms}}$ all peak near the optical surface, where radiative energy transport starts to become important (cf. Fig.~\ref{fig:strat6}). The peak rms velocity decreases with decreasing effective temperature. The position of the maximum of $\vel_{z,\mathrm{rms}}$ shifts along the model sequence: in the F- and G-star simulations this maximum is almost directly at the optical surface while in the cooler models it is about one to two pressure scale heights below it. In the deeper layers, the convection velocity decreases monotonically with increasing depth in all simulations. In the optically thin upper layers, the overshooting large-scale convective motions slow down with increasing height above the optical surface where the stratification is stable against convection. However, the $\vel_{z,\mathrm{rms}}$ drop only for about one to three scale heights, before they rise again, as shocks become more important.\par
The right panel of Figure~\ref{fig:v} shows profiles of the mean Mach number. Although the typical velocities in all simulations reach a substantial fraction of the local sound speed $c_s$, only in the atmosphere of the F3V simulation is an average Mach number of order unity reached. Surface convection is largely subsonic in our simulations of M, K, and G stars.\par
The left panel of Figure~\ref{fig:upanddown} shows the depth dependence of the relative area of the upflows plotted as functions of normalised average pressure. Below the surface layers, the upflow area is very similar in all six simulations and almost constant at about 63 to 65\,\% of the total area, which reflects the asymmetry between fast, dense downflows and slower upflows. The value for the upflow area of approximately 2/3 of the total area is in good agreement with the results of \citet{TS11} who used another code and different stellar parameters. Near the optical surface, the area fraction of the upflows drops to about 50\,\% as the strong correlation between vertical velocity and density weakens and the asymmetry between up- and downflows decreases. The low value of the relative upflow area in the upper layers of the F3V simulation of about 42 to 44\,\% can be interpreted as an effect called reversed granulation in the subadiabatic atmospheric layers \citep[][and references therein]{revgran}, which inverts the correlation between $\vel_z$ and $\varrho$. This effect is amplified by shock fronts in the F3V simulation: the material trailing the shocks which move upwards is over-dense compared to the average stratification.\par
The right panel of Figure~\ref{fig:upanddown} shows the mean speed of the upflows and the downflows. Although both speed profiles peak slightly below the optical surface in all simulated stars, the asymmetry between up- and downflows in the convectively unstable layers leads to mean downflow speeds reaching 1.6 to 1.8 times the mean upflow speed in the lower part of the box. In the atmospheres, averaged up- and downflow speeds are almost equal for the four cooler stars. In the F3V simulation, the presence of shocks leads to a steep rise of the average upflow speed in the optically thin layers, whereas the gradient of the averaged downflow speed is flatter (indicating an asymmetry reversed to the one observed below the surface).
\subsection{Temperature, pressure, and density}\label{sec:strat}
\begin{figure}
\centering
\includegraphics[width=8.5cm]{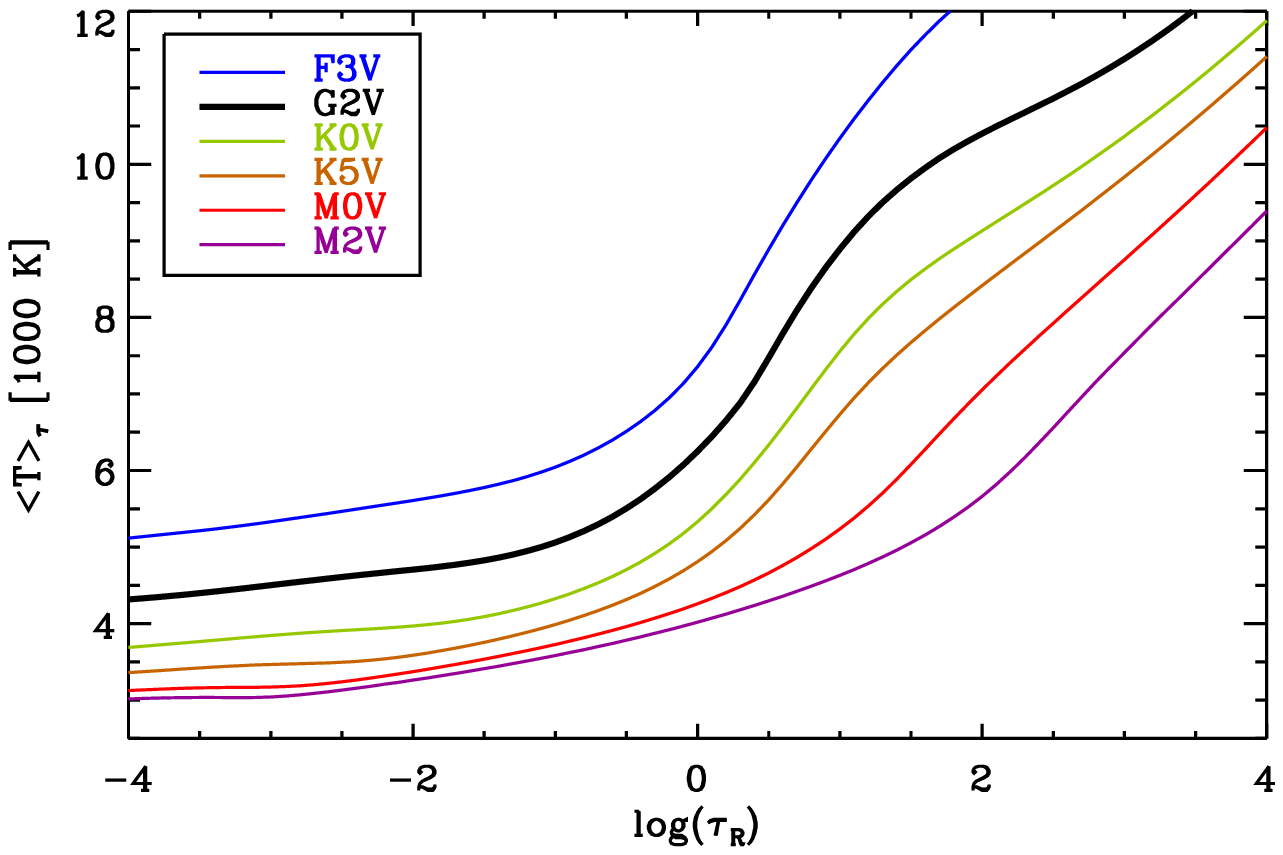}\\
\includegraphics[width=8.5cm]{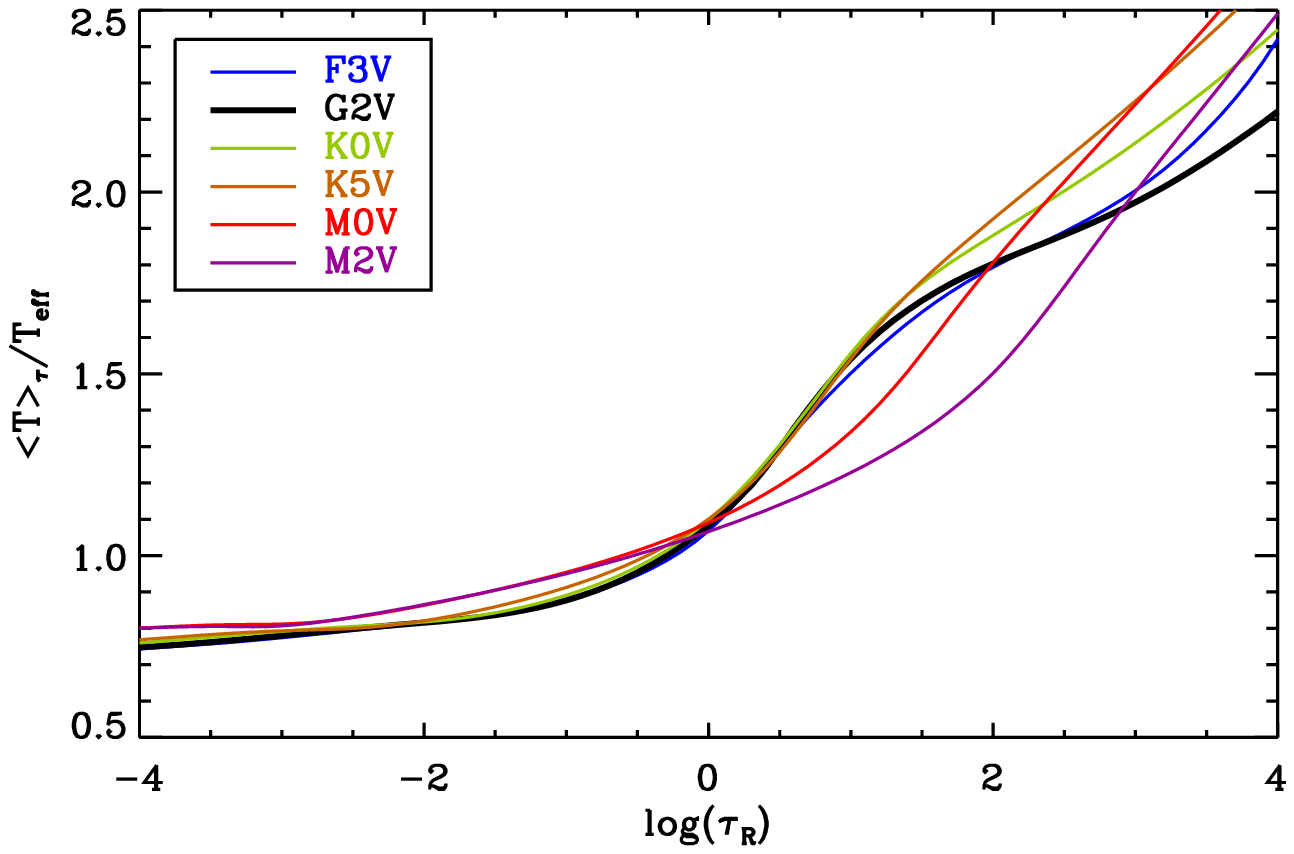}\\
\includegraphics[width=8.5cm]{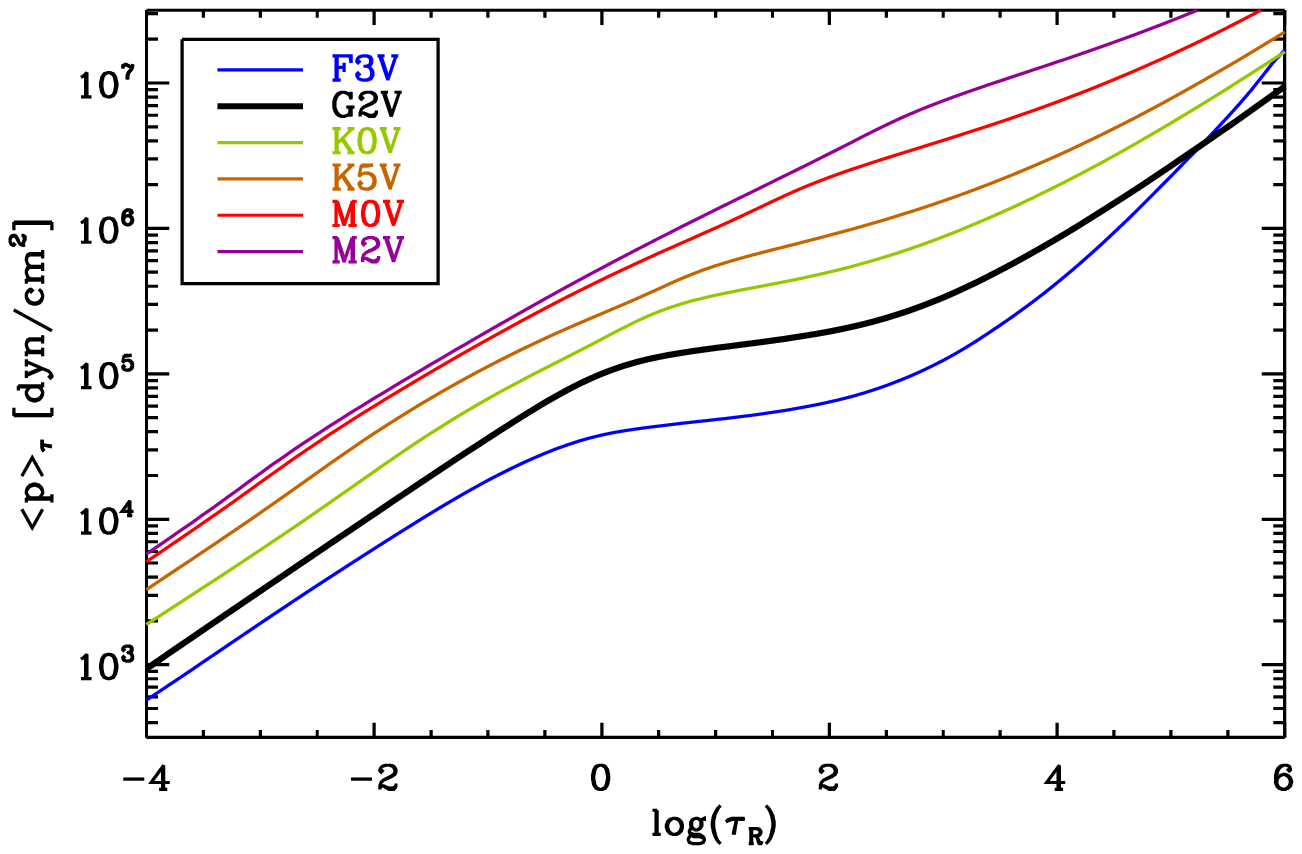}
\caption{Temperature and pressure stratifications on surfaces of constant optical depth. {\it Top panel:} temperature averaged over surfaces of constant optical depth. {\it Middle panel:} same as top panel, but normalised by the effective temperature of the respective model. {\it Bottom panel:} pressure averaged on surfaces of constant optical depth.}\label{fig:Tvontau}
\end{figure}
\begin{figure*}
\centering
\includegraphics[width=8.5cm]{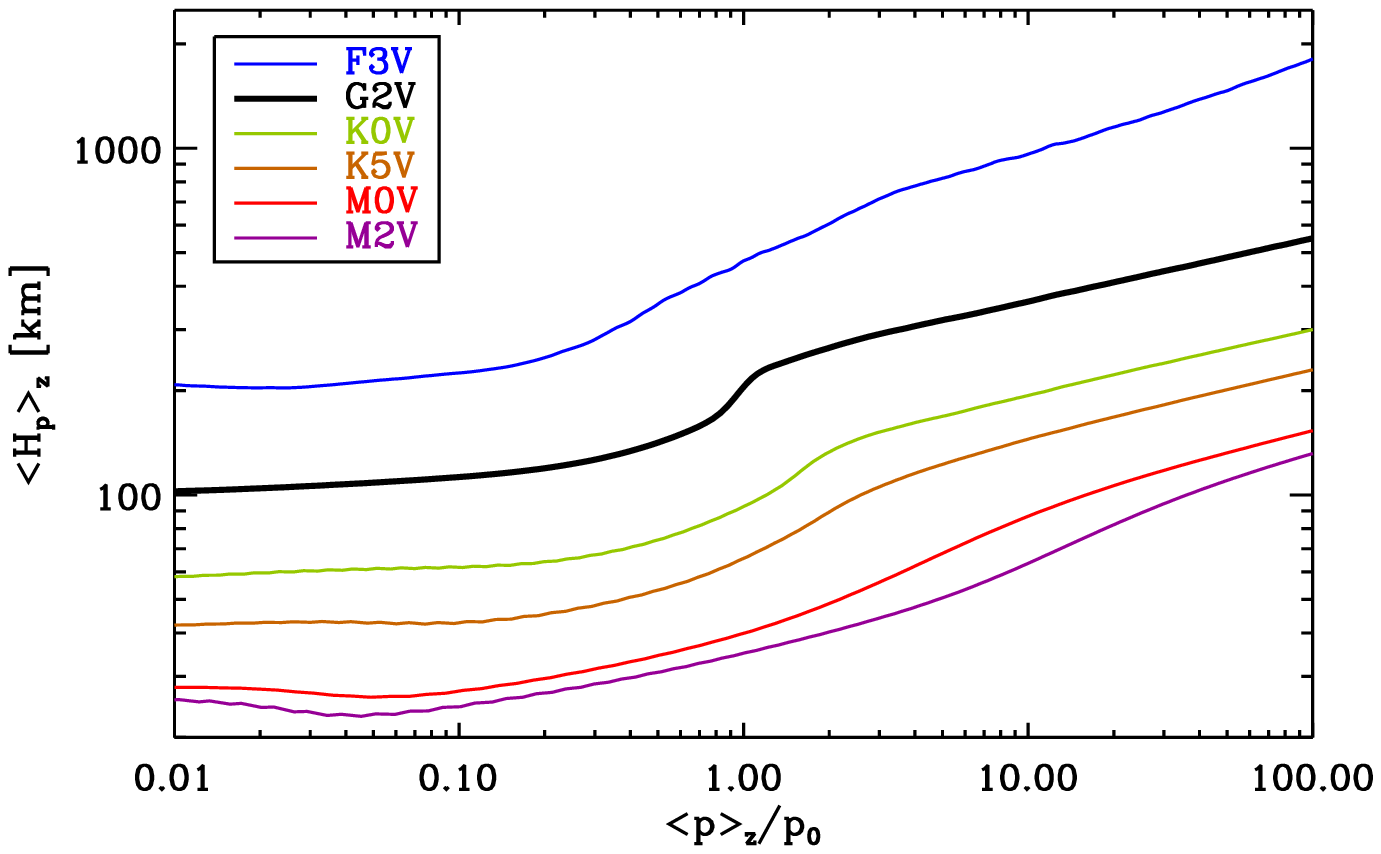}~%
\includegraphics[width=8.5cm]{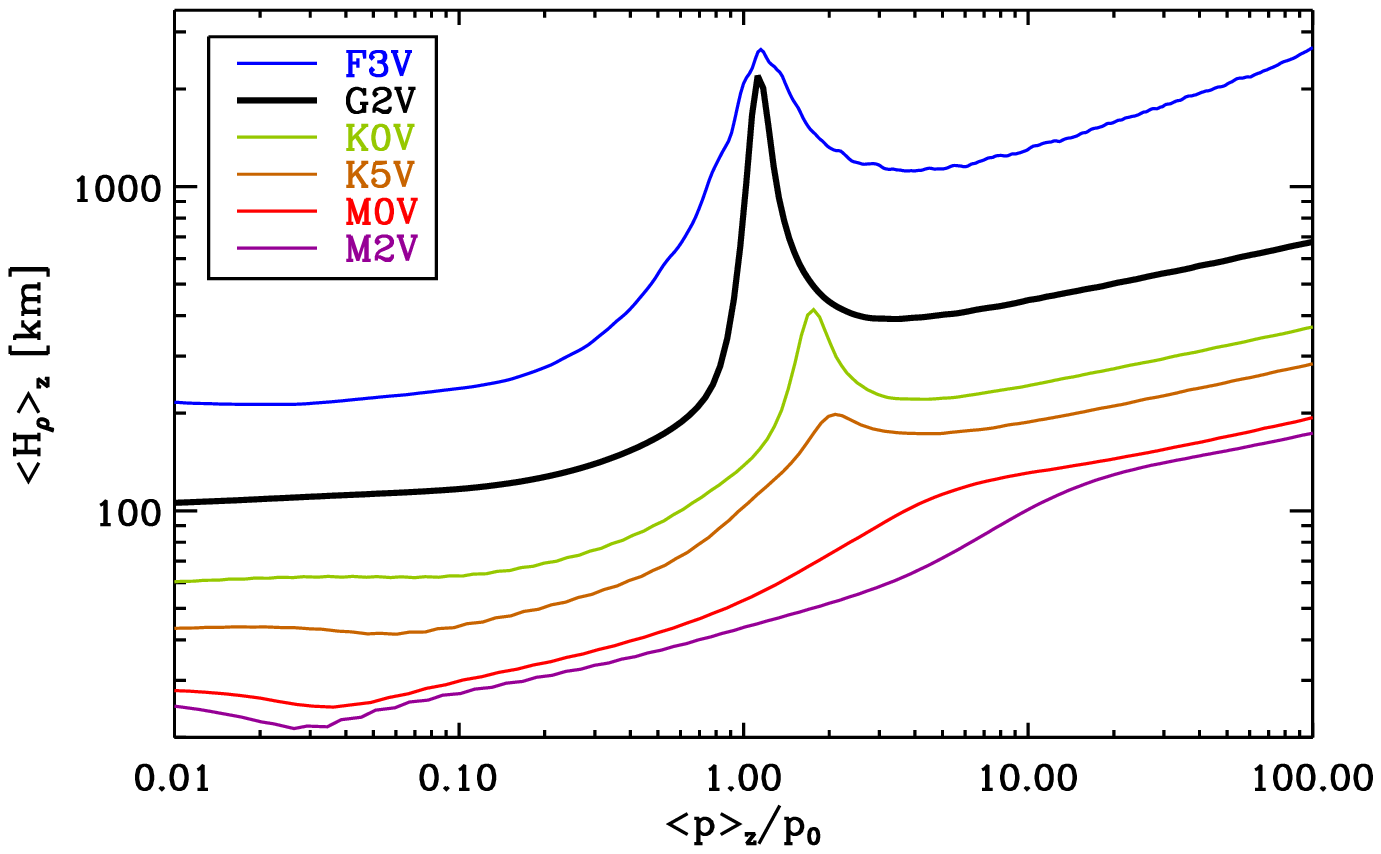}
\caption{Horizontally averaged local scale heights as functions of normalised pressure. {\it Left}: pressure scale height. {\it Right}: density scale height.}\label{fig:scaleheights}
\end{figure*}
\begin{figure*}
\centering
\includegraphics[width=8.5cm]{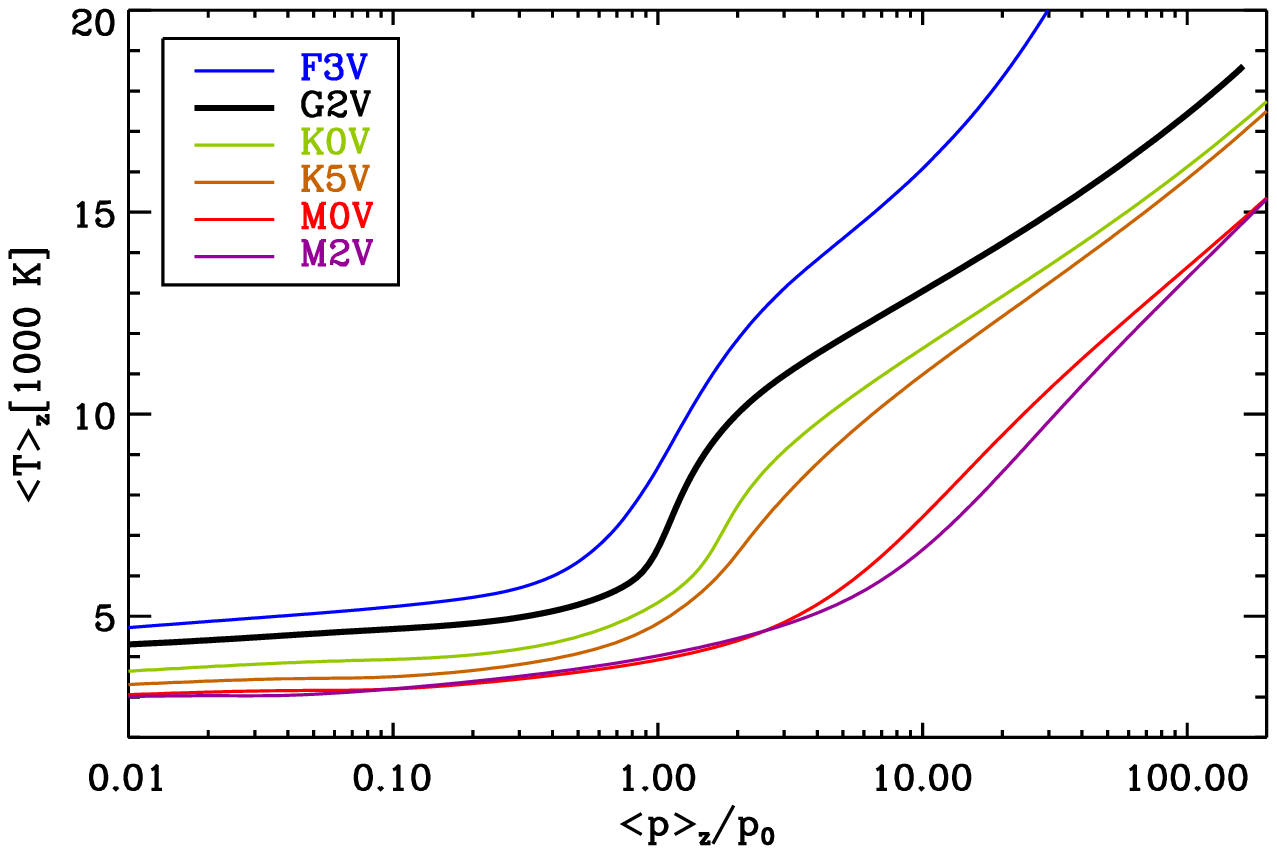}~%
\includegraphics[width=8.5cm]{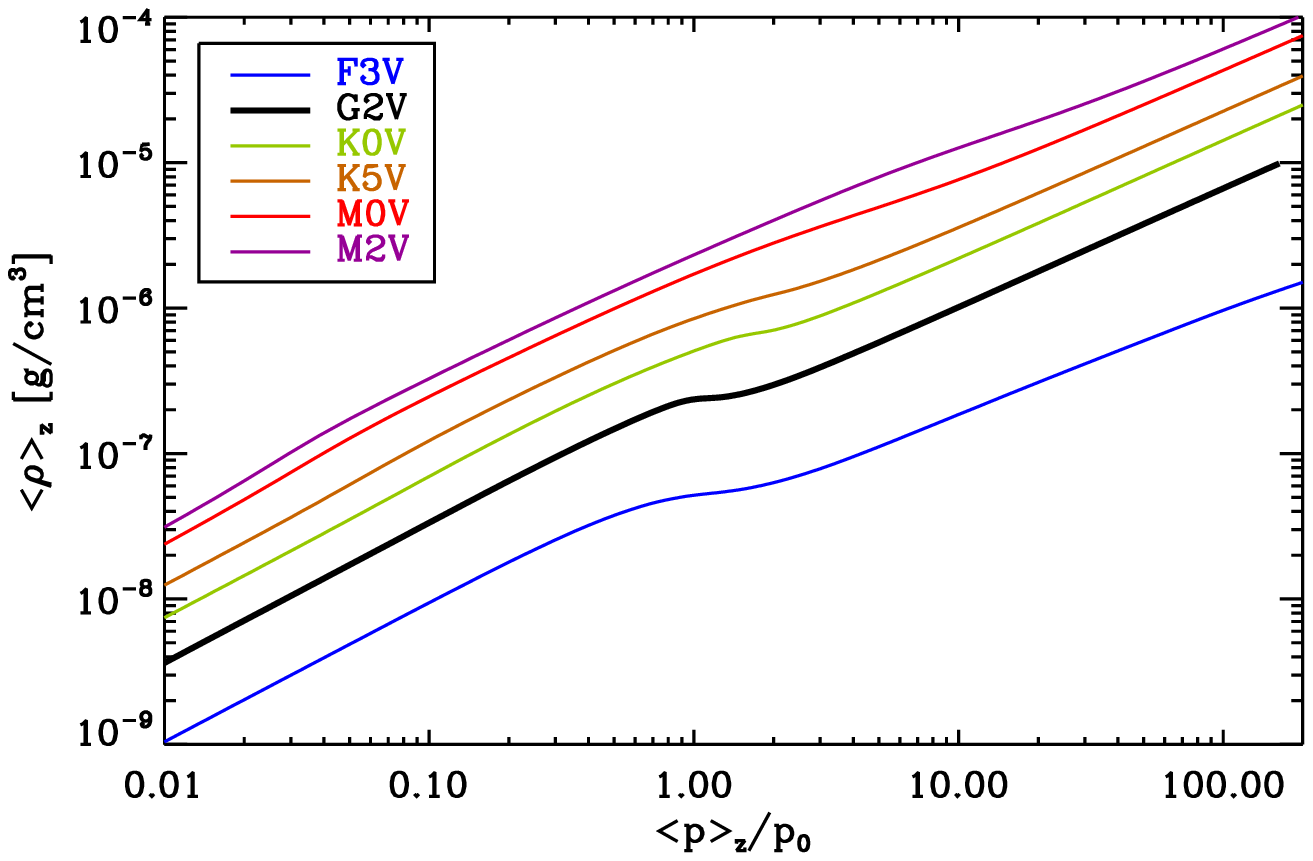}
\caption{Profiles of mean temperature $\langle T\rangle_z$ ({\it left panel}) and mean density $\langle\varrho\rangle_z$ ({\it right panel}) as functions of normalised pressure.}\label{fig:strat3}
\end{figure*}
\begin{figure*}
\centering
\includegraphics[width=8.5cm]{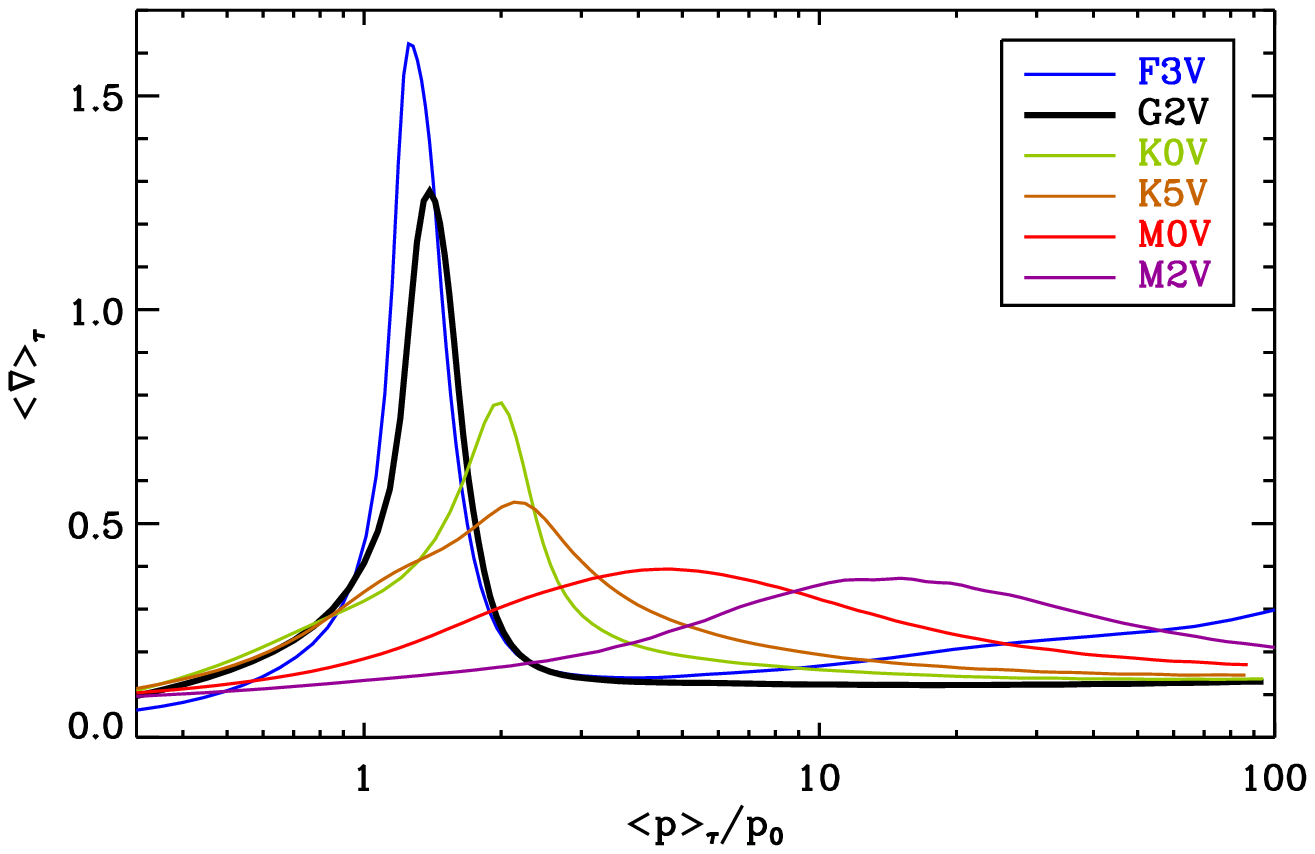}~%
\includegraphics[width=8.5cm]{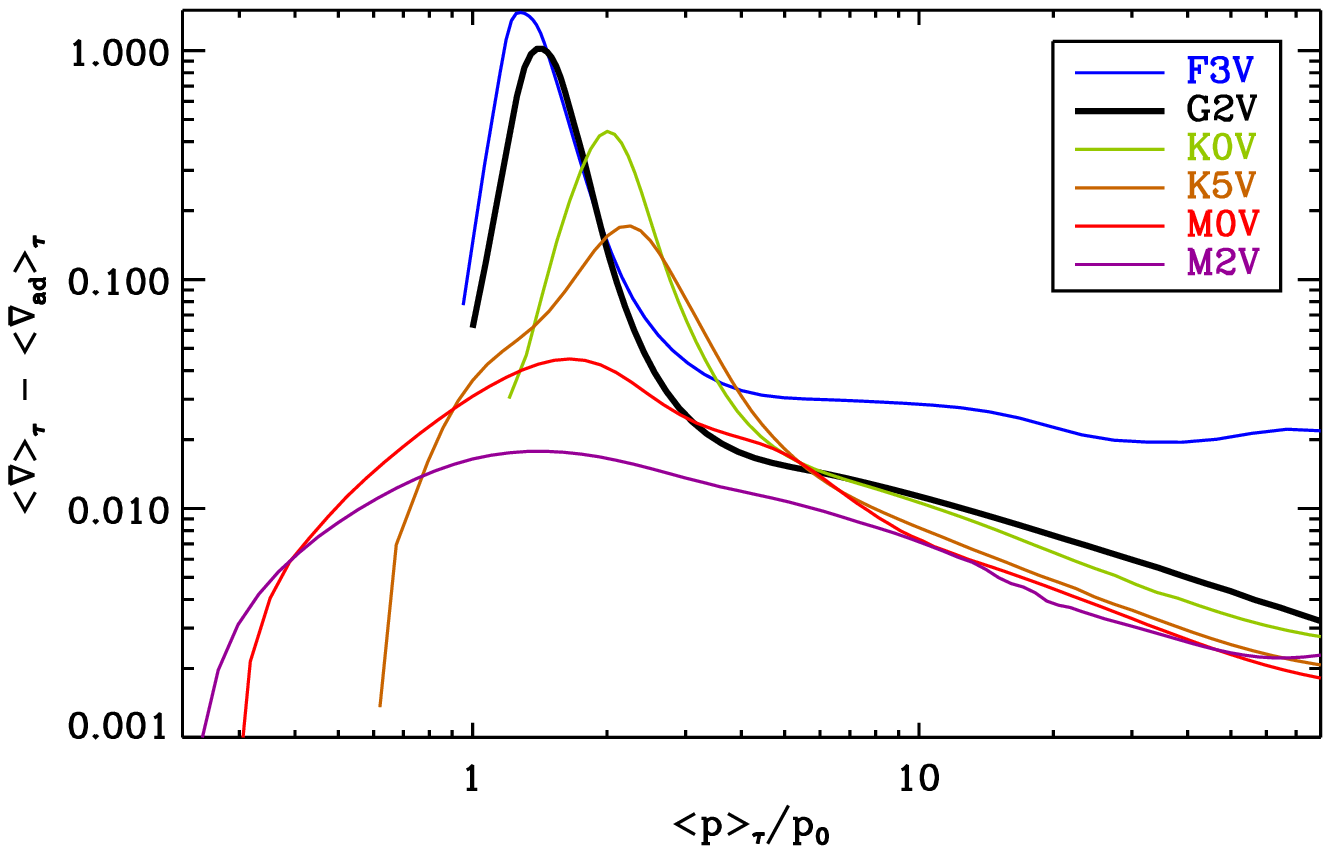}
\caption{Profiles of the logarithmic temperature gradient ({\it left panel}) and of superadiabaticity ({\it right panel}) averaged on iso-$\tau_{\mathrm{R}}$ surfaces as function of pressure.}\label{fig:strat4}
\end{figure*}
The upper two panels of Figure~\ref{fig:Tvontau} show the temperature (averaged over iso-$\tau_{\mathrm{R}}$ surfaces; for a discussion of the different averages $\langle\cdot\rangle_{\tau}$, $\langle\cdot\rangle_z$, and $\langle\cdot\rangle_p$, see Appendix~\ref{app:avg}) as function of $\log\tau_{\mathrm{R}}$, both in absolute units and normalised by $T_{\mathrm{eff}}$. The simulations from F3V to K5V show a steep temperature gradient just beneath the optical surface, whereas, in the M dwarfs, the steepest temperature gradient occurs well below this layer (see also Figs\,\ref{fig:strat3} and \ref{fig:strat4} and Sect.~\ref{sec:enfl}). In the normalised representation, all simulations have a similar profile in the atmosphere ($\log\tau_{\mathrm{R}}<0$), while their temperature curves diverge in the subphotospheric layers.\par
The bottom panel of Figure~\ref{fig:Tvontau} gives the pressure (averaged on iso-$\tau_{\mathrm{R}}$ surfaces) plotted as function of $\log\tau_{\mathrm{R}}$. In the atmosphere, where opacity and temperature are only mildly height-dependent, $\log p$ essentially depends linearly on $\log\tau_{\mathrm{R}}$ as the structure is governed by hydrostatic and radiative equilibrium and is also almost iso-thermal. In the layers just below the photosphere, the curves for the different simulations diverge. The diverging profiles of the subphotospheric temperature (middle panel) and pressure (bottom panel) reflect that the pressure and temperature structures are determined by convection below the photosphere and converge to different adiabat (depending on the stellar parameters) in the deep convective envelopes.\par
Figure~\ref{fig:scaleheights} shows the depth dependences of the pressure and density scale heights as functions of normalised pressure. We define horizontally averaged local scale heights as
\begin{equation}\label{def:Hrho}
\langle H_{p}\rangle :=\left(\mathrm{d}\log\langle p\rangle_z/\mathrm{d}z\right)^{-1}\quad\mathrm{and}\quad\langle H_{\varrho}\rangle :=\left(\mathrm{d}\log\langle\varrho\rangle_z/\mathrm{d}z\right)^{-1}\,\,.
\end{equation}
 As the gravitational acceleration increases and the photospheric temperature decreases monotonically from F3V to M2V the, local pressure scale height around $\tau_{\mathrm{R}}=1$ decreases from $\sim 500\,\mathrm{km}$ in F3V to $\sim 35\,\mathrm{km}$ in M2V. In the atmosphere, where the temperature is mildly height-dependent, the local pressure scale height becomes roughly constant. In the convective layers, the strong temperature gradient entails also a strong increase of the pressure scale height towards deeper layers. At $p=100\,p_0$, near the bottom of the simulation boxes, the local pressure scale height of the F3V simulation is already $\sim 2000\,\mathrm{km}$, which poses a problem for the current implementation of the \texttt{MURaM} code with its fixed vertical cell size (high computational costs).\par
The peak of the density scale height near the optical surface in some simulations coincides with the strong photospheric temperature gradient of these simulations. Locally, the density scale height often becomes negative at the optical surface in the F3V and G2V simulations (density inversion). In the subsurface layers with high temperature gradient, the density scale heights are somewhat (M2V-G2V:15-30\%, F3V: up to 45\%) larger than the pressure scale heights whereas, in the almost isothermal atmospheres, the scale heights of pressure and density are almost equal.\par
Figure\,\ref{fig:strat3} shows the profiles of temperature and density averaged on iso-$z$ surfaces as functions of the normalised pressure. As already seen in Fig.~\ref{fig:Tvontau} on the $\tau_{\mathrm{R}}$ scale, the coolest models lack the strong photospheric temperature gradient of the warmer models. This is shown quantitatively in the left panel of Figure~\ref{fig:strat4}, which gives the mean profiles of the logarithmic temperature gradient, $\langle\nabla\rangle_{\tau}=\mathrm{d}\log \langle T\rangle_{\tau} / \mathrm{d}\log \langle p\rangle_{\tau}$. Here, the iso-$\tau$ average was chosen because $\nabla$ changes considerably near the optical surface. Since this represents a transition from a (highly) superadiabatic to a subadiabatic regime, averaging over iso-$z$ planes would smear out the sharp photospheric feature in the temperature gradient and thus obscure the relevant physics in this layer.\par
In the right panel of Figure~\ref{fig:strat4}, the profile of the superadiabaticity $\langle\nabla\rangle_{\tau}-\langle\nabla_{\mathrm{ad}}\rangle_{\tau}$ is given. The superadiabaticity in the lowest part of the simulation domain is small (${\sim}10^{-3}$) for most of the models, with the exception of the F3V simulation, where the stratification remains substantially superadiabatic even 5 pressure scale heights below the optical surface. For the M dwarfs, the superadiabaticity is low compared to the hotter models, even in the layers directly beneath the optical surface. This is a consequence of the high densities (i.\,e. high heat capacity per volume) and low energy fluxes and consequently low horizontal temperature fluctuations.\par
\begin{figure}
\centering
\includegraphics[width=8.5cm]{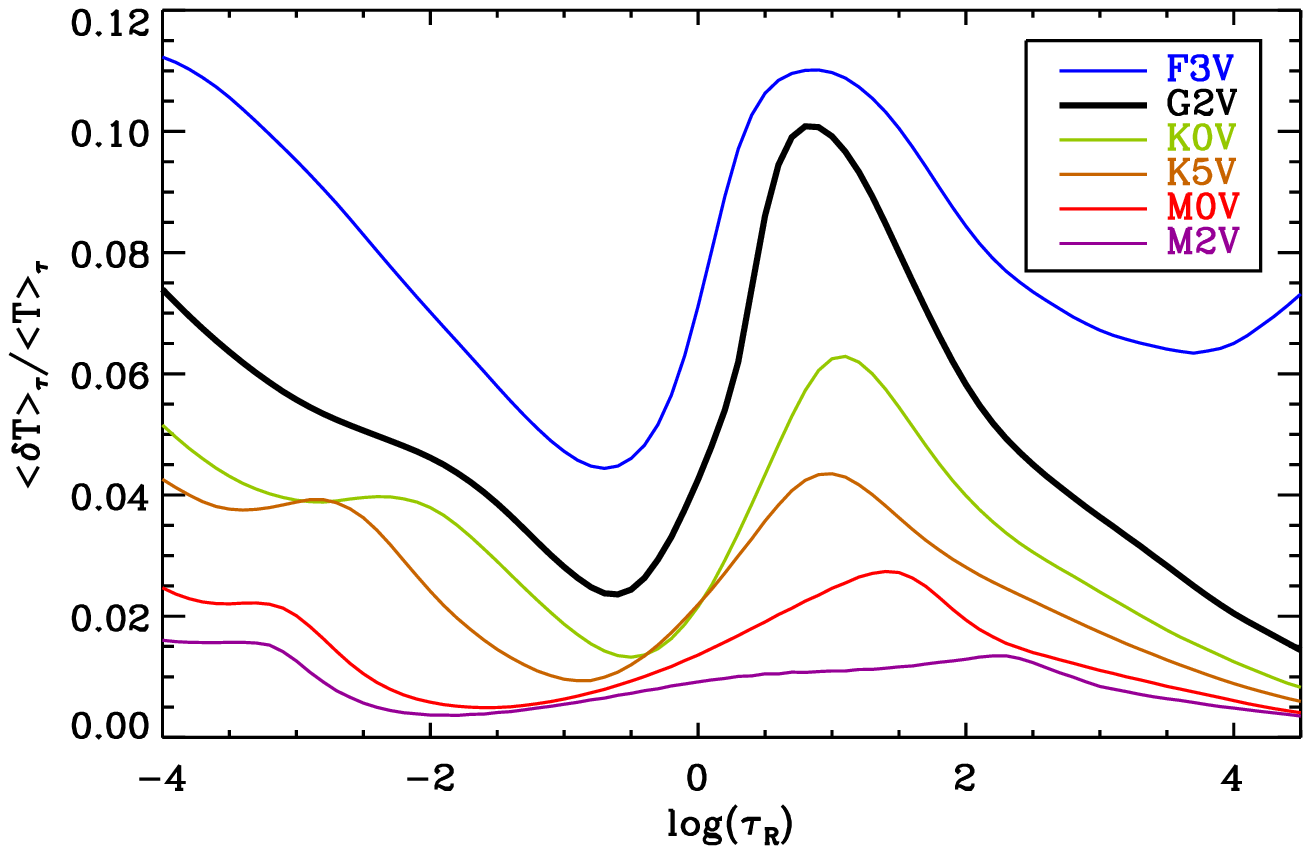}\\
\includegraphics[width=8.5cm]{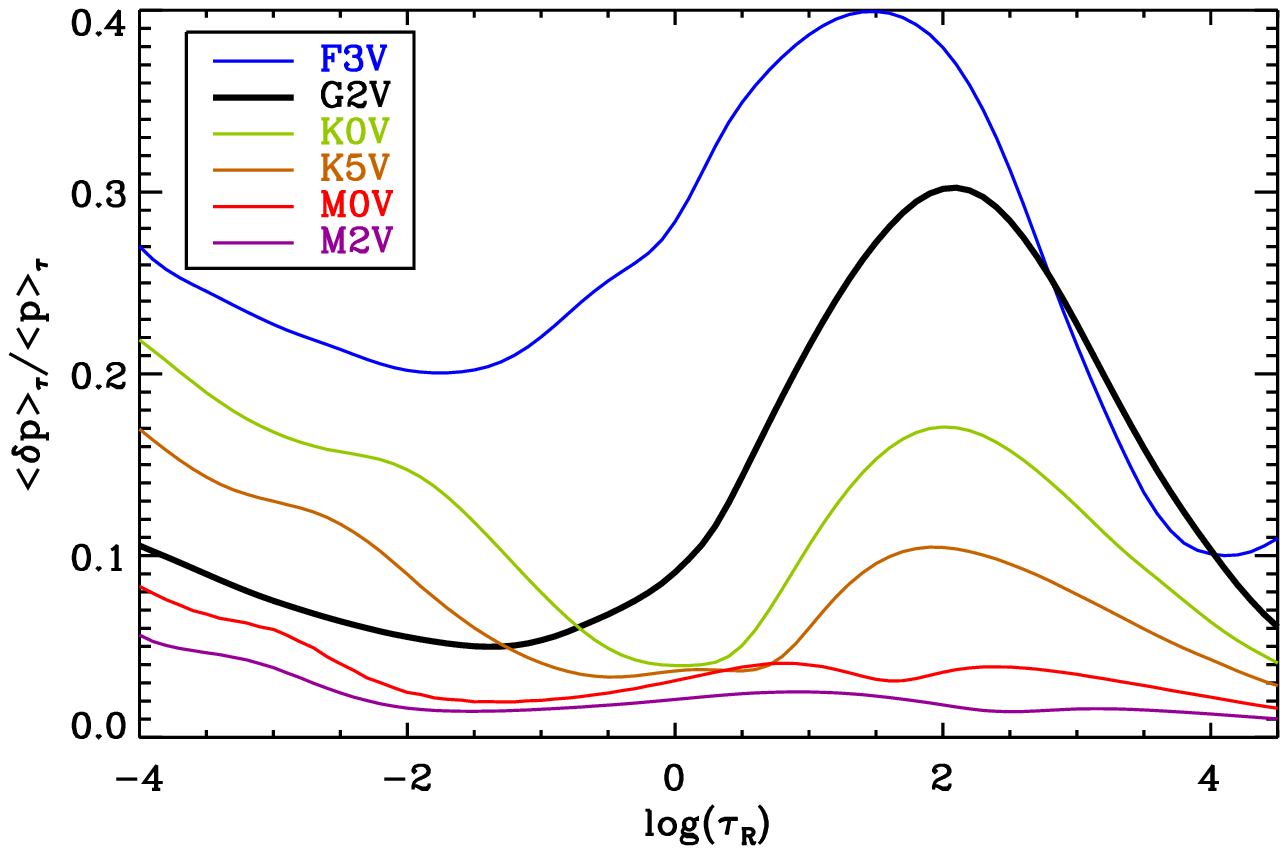}\\
\includegraphics[width=8.5cm]{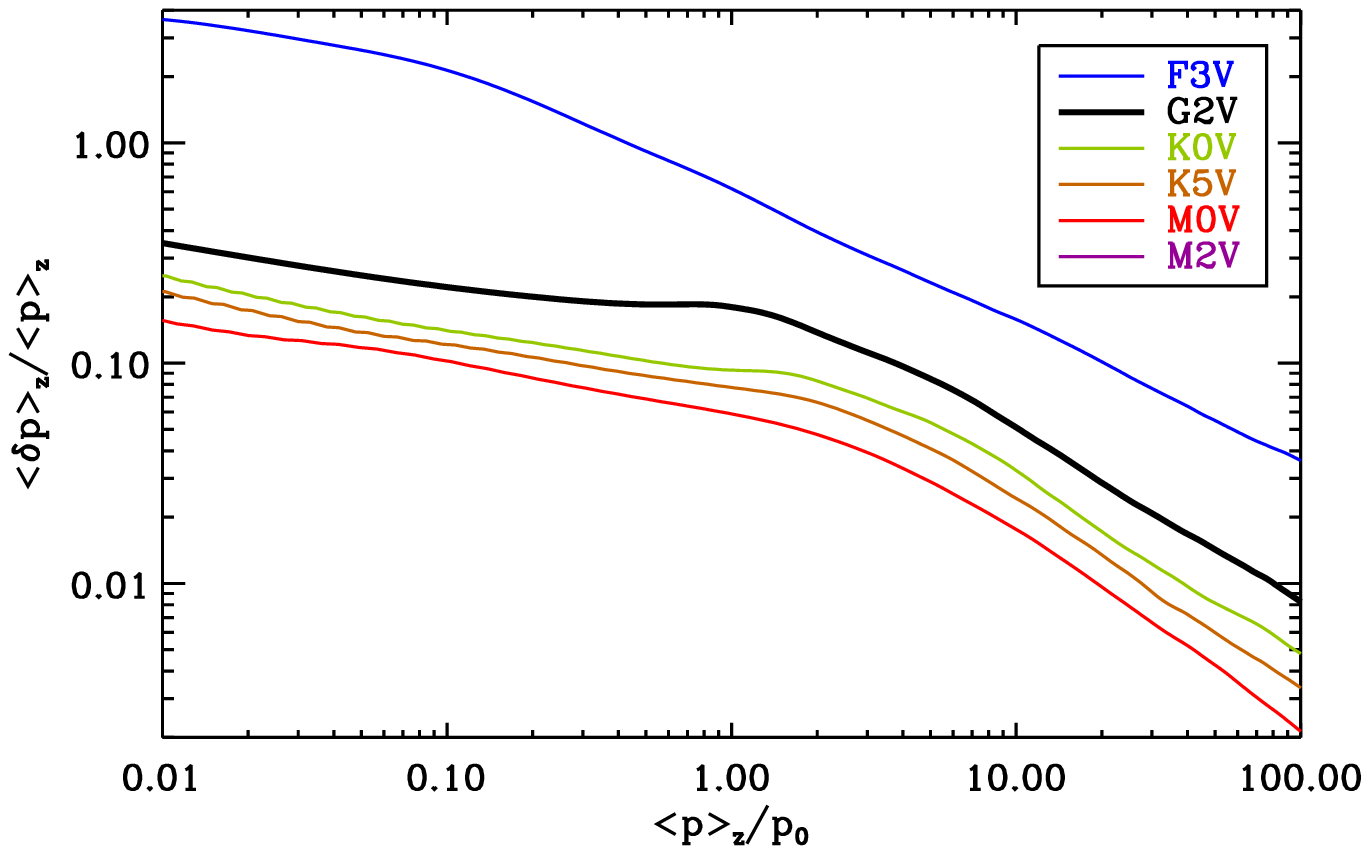}\\
\caption{Profiles of relative rms fluctuations. {\it Top}: temperature fluctuations on surfaces of constant optical depth. {\it Middle}: gas pressure fluctuations on surfaces of constant optical depth. {\it Bottom}: pressure fluctuations on surfaces of constant geometrical depth.}\label{fig:strat5}
\end{figure}
The top panel of Figure~\ref{fig:strat5} shows the relative rms fluctuations of temperature on surfaces of constant optical depth. The relative rms fluctuations of temperature show a monotonic decrease from the hotter to the cooler stars at all depths. This is consistent with the trend in the bolometric intensity contrast (cf. Table~\ref{tab:values}). The subphotospheric peak in the depth profile of temperature fluctuations is at lower optical depth in the F- and G-star simulations compared to the cooler simulations. This has already been pointed out by \citet{ND90a}, who coined the expression ``hidden'' or ``veiled'' granulation for stars cooler than the Sun, as the maximum temperature contrast occurs far below the optical surface. In the case of the K0V star, the relative temperature contrast at the optical surface is only about 34\% of its peak value at $\log\tau_{\mathrm{R}}\approx 1$ (compared to 42\% in the solar simulation and 64\% in the F3V simulation). The reason for this effect is the lower temperature-sensitivity of opacity near the optical surfaces of the cooler K and M stars. This leads to the transition from convective to radiative energy transport occuring at somewhat larger optical depth or normalised pressure (particularly in the K-star simulations) and over a larger optical depth range or normalised pressure range (particularly in the M-star simulations; see Sect.~\ref{sec:enfl}).\par
The middle panel of Figure~\ref{fig:strat5} shows the rms fluctuations of gas pressure on surfaces of constant optical depth. They also diminish with decreasing effective temperature of the simulations, with the notable exception of the G-type star, where the rms fluctuations of pressure on surfaces of constant optical depth are very low in the upper atmosphere compared to the much cooler K stars. This can be explained as an opacity effect: while the temperature dependence of the Rosseland opacity $\kappa_{\mathrm{R}}(p,T)$ is usually much more important than the pressure dependence, in the temperature range between 4000 and $5000\,\mathrm{K}$ and at pressures between $10^2$ and $10^5\,\mathrm{dyn\,cm^{-2}}$, $\kappa_{\mathrm{R}}$ is nearly independent of temperature. Moreover, the temperature fluctuations in the atmosphere are in general relatively small, so that density depends mainly on pressure. Therefore, the increment of the optical depth $\mathrm{d}\tau_{\mathrm{R}}=\left(\kappa_{\mathrm{R}}\varrho\right)\mathrm{d}z$ becomes (almost) independent of temperature, too, so that iso-$\tau_{\mathrm{R}}$ and iso-$p$ surfaces of an atmosphere in this temperature regime are almost identical, i.\,e. the fluctuations of pressure on surfaces of constant optical depth become small. In our simulation sequence, the G2V star is the only star where the temperature and pressure of the atmospheric layers at $-4\le \log\tau_{\mathrm{R}}\le -1$ fall in the regime of nearly temperature-independent $\kappa_{\mathrm{R}}$.\par
The bottom panel of Figure~\ref{fig:strat5} shows the rms fluctuations of pressure on planes of constant geometrical depth, which are mainly monotonically decreasing with increasing depth and with decreasing effective temperature. This illustrates that in the deeper layers, where velocities and horizontal temperature fluctuations are small, the deviations from hydrostatic equilibrium are small, too. Horizontal temperature fluctuations entail horizontally varying local pressure scale heights: the pressure fluctuations on a horizontal plane in the atmosphere can be regarded as the integrated effect of the temperature fluctuations below this plane.\\
Before the energy balance of the simulated stellar surface layers is analysed in Sect.~\ref{sec:enfl}, we want to point out that the trends observed in convective velocities and temperature fluctuations can be consistently explained as an effect of the stellar parameters: the lower temperature and higher gravitational acceleration in the atmospheres of cooler stars result in much higher densities. The resulting higher heat capacity per unit volume, $c_p/\varrho$, and the much lower net energy flux ($F\propto T_{\mathrm{eff}}^4$) then have the consequence that the convective motions in the coolest models are less vigorous. This is reflected in the rms velocities of about $0.5\,\mathrm{km\,s^{-1}}$ for those simulations compared to more than $5\,\mathrm{km\,s^{-1}}$ for the F3V-star simulation (see Fig.~\ref{fig:v}). A rough estimate of the convective heat flux $F_{\mathrm{conv}}$ (enthalpy flux) in the spirit of mixing-length theory gives
\begin{equation}
F_{\mathrm{conv}}\approx \Delta T\cdot \vel_{\mathrm{conv}}\cdot c_p \cdot \varrho\,,
\end{equation}
where $\Delta T$ is the temperature contrast between up and downflows, $\vel_{\mathrm{conv}}$ is the convective velocity, and $\varrho$ is the density, both of which are assumed to be equal in up- and downflows for this rough estimate. In the convective subsurface layers of cool stars radiative energy transport can be neglected, so that 
\[
F_{\mathrm{conv}}\approx F_{\mathrm{tot}}=\sigma T_{\mathrm{eff}}^4\,.
\]
If we approximate the quantities $\vel_{\mathrm{conv}}$ and $\Delta T$ by $\langle \vel_{z,\mathrm{rms}}\rangle$ and $\langle \delta T\rangle$, respectively ( $\langle \cdot \rangle$ denotes a horizontal and temporal average), and if we further ignore the variation of $c_p$  with temperature due to ionisation, we obtain:
\begin{equation}\label{eqn:conv}
T_{\mathrm{eff}}^4\langle\varrho\rangle^{-1}\sim \langle\delta T\rangle \cdot \langle \vel_{z,\mathrm{rms}}\rangle\,.
\end{equation}
Directly beneath the optical surface, the product on the left-hand side decreases by a factor of about 500 from F3V to M2V, which entails a strong variation of $\langle \delta T\rangle\cdot\langle \vel_{z,\mathrm{rms}}(z_0)\rangle$. In fact, the simulations show that both $\langle\delta T\rangle$ and $\langle \vel_{z,\mathrm{rms}}(z_0)\rangle$ decrease monotonically through the model sequence by about an order of magnitude each (see Figs~\ref{fig:Dhor} and~\ref{fig:strat5}).\\ 

\subsection{Energy flux}\label{sec:enfl}
The total energy flux leaving the stellar atmosphere in the form of radiation is supplied by energy flux from below. In the absence of nuclear energy sources, the (temporally averaged) total luminosity is constant throughout the upper layers of a star. In plane-parallel geometry, this means that the energy flux is independent of depth.\par 
In the convection zone, the energy flux is almost entirely provided by convective energy transport and is mainly composed of two opposing fluxes: the net enthalpy flux which is directed towards the surface, and the net kinetic energy flux, which is directed inwards. This is a consequence of the asymmetry in temperature and velocity. The fast, cool, and dense downdrafts cary less enthalpy but more kinetic energy (per unit area) than the slow, hot upflows. The left panel of Figure~\ref{fig:strat6} shows the profiles of these two fluxes for the six simulated stars, all normalised to the respective total flux. Most models show profiles of the kinetic energy flux levelling off below the optical surface at values between $\sim$20 to $\sim$60\,\% of the total flux, larger values corresponding to cooler stars. The F3V model, however, shows a monotonic increase of the kinetic energy flux for increasing depth. 
The right panel of Figure~\ref{fig:strat6} shows the radiative flux in the simulations. The transition from purely convective ($F_{\mathrm{rad}}/F_{\mathrm{tot}}\lesssim 0.1$) to mainly radiative energy transport ($F_{\mathrm{rad}}/F_{\mathrm{tot}} \gtrsim 0.9$) is quite sharp (within one pressure scale height), except for the M-star simulations, for which this transition takes place over a more extended pressure range. For the K- and M-star simulations, the contribution of the radiative flux to the total flux is larger within the first few pressure scale heights below the surface. One pressure scale height below the surface (i.\,e. at $p/p_0=2.72$), in the M0V model, radiation carries already 17\,\% of the total flux compared to 1.2\,\% in the G2V simulation. Half a pressure scale height below the optical surface ($p/p_0=1.65$), radiation caries 60 to 70\,\% of the flux in the K-star simulations, about 25 to 30\,\% in the M-star models, but only 13.4\,\% in the simulation of the G2V star. This means that the convective flux directly below the optical surface drops to a relatively small fraction of the total flux for the cooler simulations (especially the K-star simulations). As the convective flux is linked to vertical velocities and temperature contrast between up- and downflows, this is also reflected in the depth dependence of the rms of the vertical flow velocity (see Fig.~\ref{fig:v}) and the depth profile of the relative rms fluctuations of temperature (see top panel of Fig.~\ref{fig:strat5}). The peak of the vertical velocity rms is almost exactly at $\tau_{\mathrm{R}}=1$ in the F- and G-star simulations but somewhat deeper in the simulations of cooler stars. Similarly, the peak of the relative temperature fluctuations shifts to higher optical depth from hotter to cooler models. Near the bottom of the simulation box, the contribution of the radiative flux eventually becomes negligible in all simulations except for F3V, where even at this depth radiation still carries about 0.5 \% of the energy flux. The much lower density and higher temperature in the subsurface layers of this star enable a somewhat more efficient radiative heat transport.\par
The inset in the right panel of Figure~\ref{fig:strat6} indicates the negative convective energy flux (which follows directly from $\langle F_{\mathrm{rad}}\rangle_z > F_{\mathrm{tot}}$) in the convective overshoot region, which is most prominent in the K-dwarf models.\par
\begin{figure*}
\centering
\includegraphics[width=8.5cm]{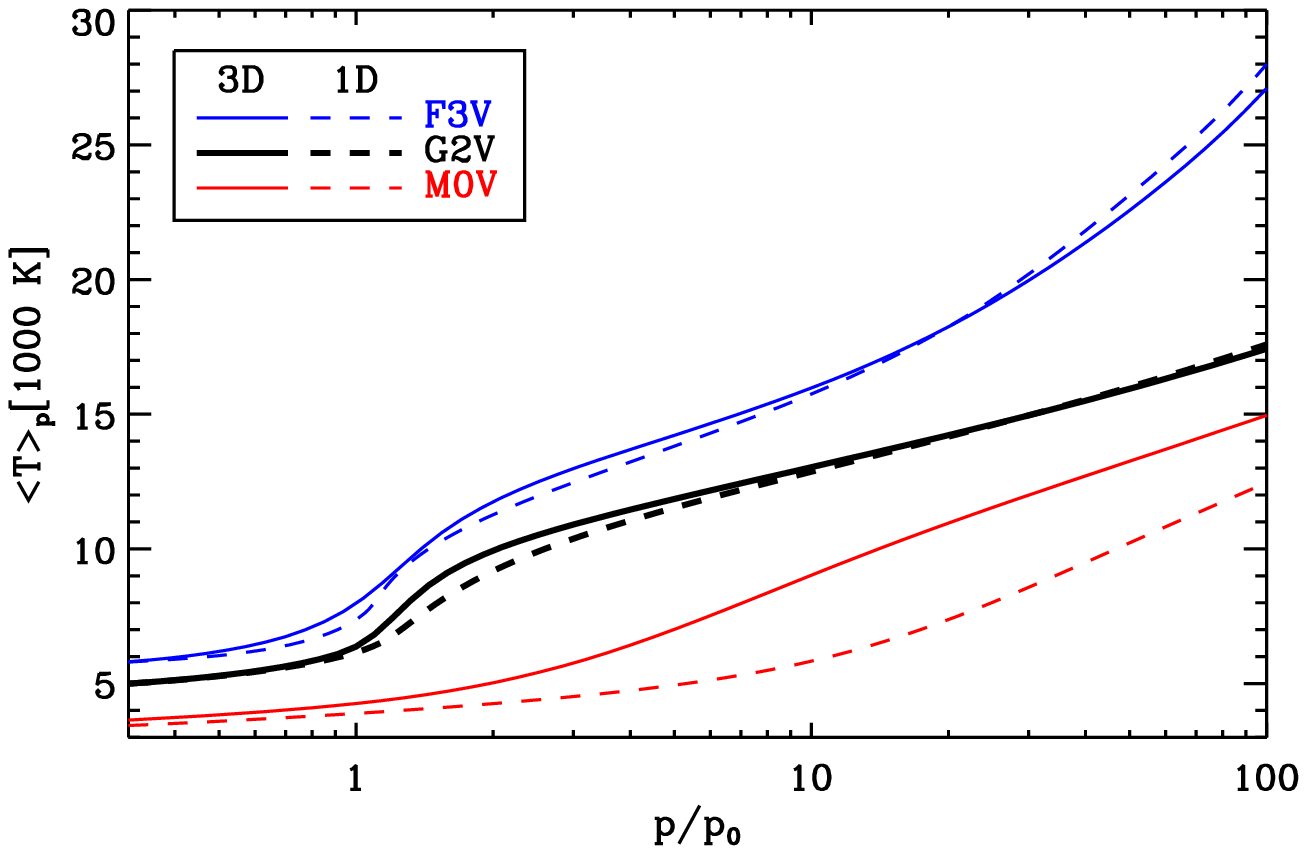}~%
\includegraphics[width=8.5cm]{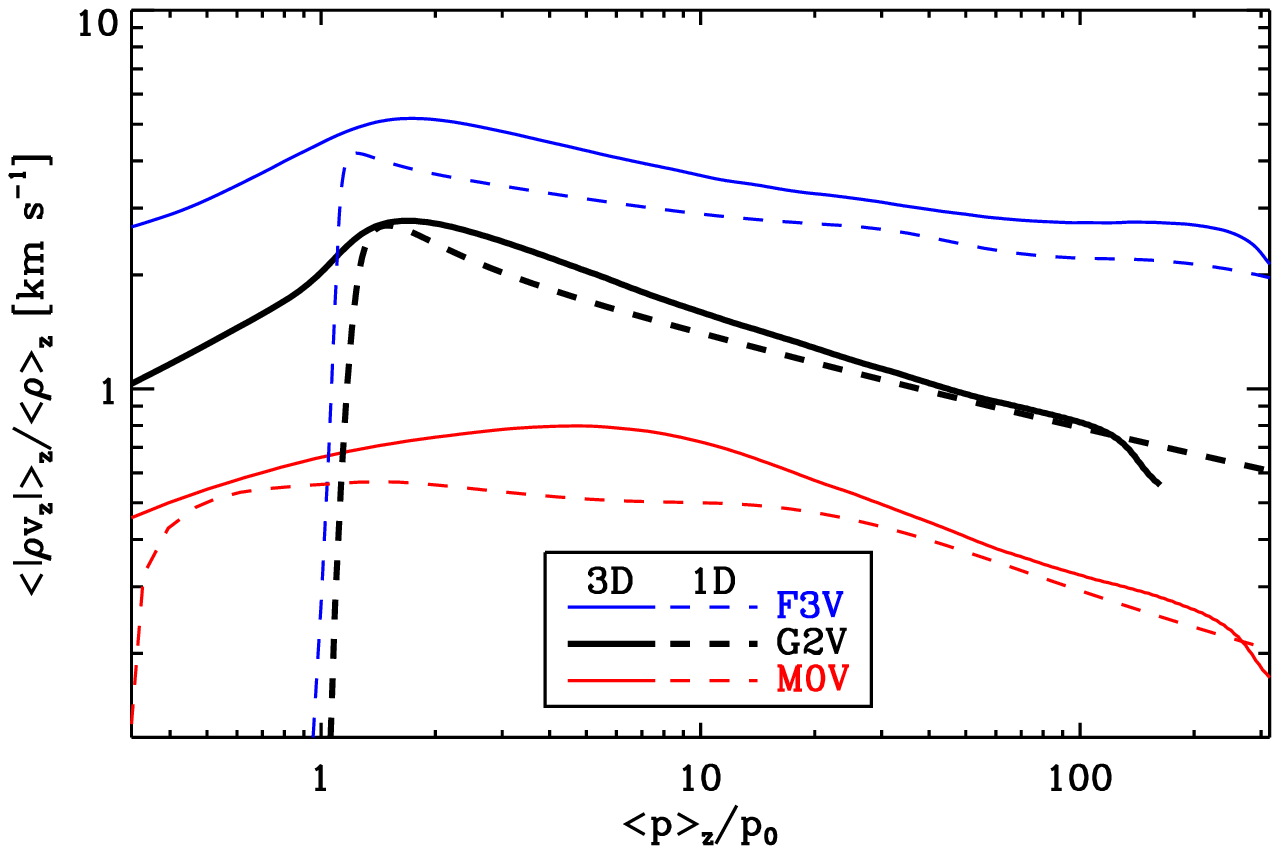}\\
\caption{Comparison between the 3D \texttt{MURaM} simulations with 1D MLT models. The dashed curves represent results from a 1D MLT model atmosphere and the solid curves represent the averaged simulation results. {\it Left}: Profile of temperature (3D: averaged on iso-$p$ surfaces). {\it Right}: average vertical velocity, weighted by density (3D: averaged on iso-$z$ surfaces; 1D: convective velocity). In the 1D models, the mixing-length parameter $\alpha$ was set to 1.5, 1.7, and 2.0 for F3V, G2V, and M0V, respectively.}\label{fig:1Dvs3D1}
\end{figure*}
\subsection{Comparison to 1D models}
The horizontal averages of our 3D models can be compared to 1D mixing-length models. We used a model grid by Ludwig (priv. comm.) of 1D mixing-length calculations. Analogous to the definition of $\langle z\rangle_{\tau_{\mathrm{R}}=1}\equiv 0$ and $\langle p\rangle_{\tau_{\mathrm{R}}=1}\equiv p_0$ as reference points for the $z$- and $p$-scales in the averaged 3D stratifications, we chose $z(\tau_{\mathrm{R}}=1)\equiv 0$ and $p(\tau_{\mathrm{R}}=1)\equiv p_0$ as reference points for the 1D models.\par
Figure~\ref{fig:1Dvs3D1} shows depth profiles of temperature and rms of the vertical velocity as functions of pressure for three of our simulations (F3V, G2V, and M0V) and 1D models with the same $g$ and $T_{\mathrm{eff}}$. The mixing-length parameter was set to $\alpha=1.5$, 1.7, and 2.0 for F3V, G2V, and M0V, respectively. Although the temperature profiles in the almost adiabatic subsurface layers are influenced by the choice of $\alpha$, they could not be brought into exact agreement with the profiles of the 3D results (which, moreover, depend on the averaging method). We therefore chose $\alpha$ close to the values by \citet{TS11} for the mass mixing length, which are consistent with the literature values for the MLT-$\alpha$ cited there.\par
The left panel of Figure~\ref{fig:1Dvs3D1} shows the run of temperature. The 3D results were averaged on surfaces of constant pressure (as a compromise between the iso-$z$ and iso-$\tau_{\mathrm{R}}$ averages, see Appendix~\ref{app:avg}). The general shape of the curves does not differ strongly between averaged 3D and 1D results. However, the temperatures in the 1D models are in general somewhat lower than in the averaged 3D simulations, which can be partly explained by the way in which the 3D results were horizontally averaged but can also be an effect of the opacity in the optically thin layers, which differs between 1D and our 3D models \citep[the 1D models were calculated with \texttt{ATLAS6} opacities, see][]{ATLAS6}. The position and steepness of the strong photospheric gradient does not match the result of the more realistic 3D simulations. This has two reasons: First, this feature is very sensitive to the horizontal averaging method (see Appendix~\ref{app:avg}) because of the strongly corrugated optical surfaces. Second, in this layer overshoot and the transition from convective to radiative energy transport play a major role. The physics behind these effects is essentially three-dimensional and has to be parameterised in a 1D model where only a very crude description of these effects is possible. In the lower part of the depth range considered, there is also a mismatch between the temperature gradients of the averaged 3D and the 1D results. A disparity in the superadiabaticity (particularly for F3V) and differences in the equation of state between the 3D and 1D models are responsible for this deviation.\par
In the right panel of Figure~\ref{fig:1Dvs3D1}, we show the run of the vertical velocity. For the 3D results, iso-$z$ averages of the density-weighted vertical flow speed are shown. While the gradient of the subsurface velocity and the position of the velocity peak (at least for F3V and G2V) are similar between 1D and 3D models, the 1D models have lower velocities than the 3D simulations (by 10-30\%) in the nearly adiabatic interior. The 1D models obviously lack any velocities in the convectively stable layers, where the simulations display overshooting flows.\par
\begin{figure}
\centering
\includegraphics[width=2.95cm]{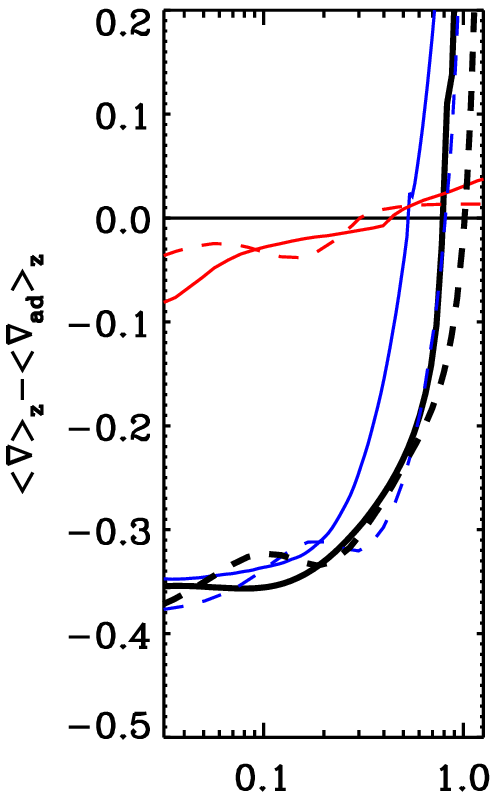}\includegraphics[width=5.55cm]{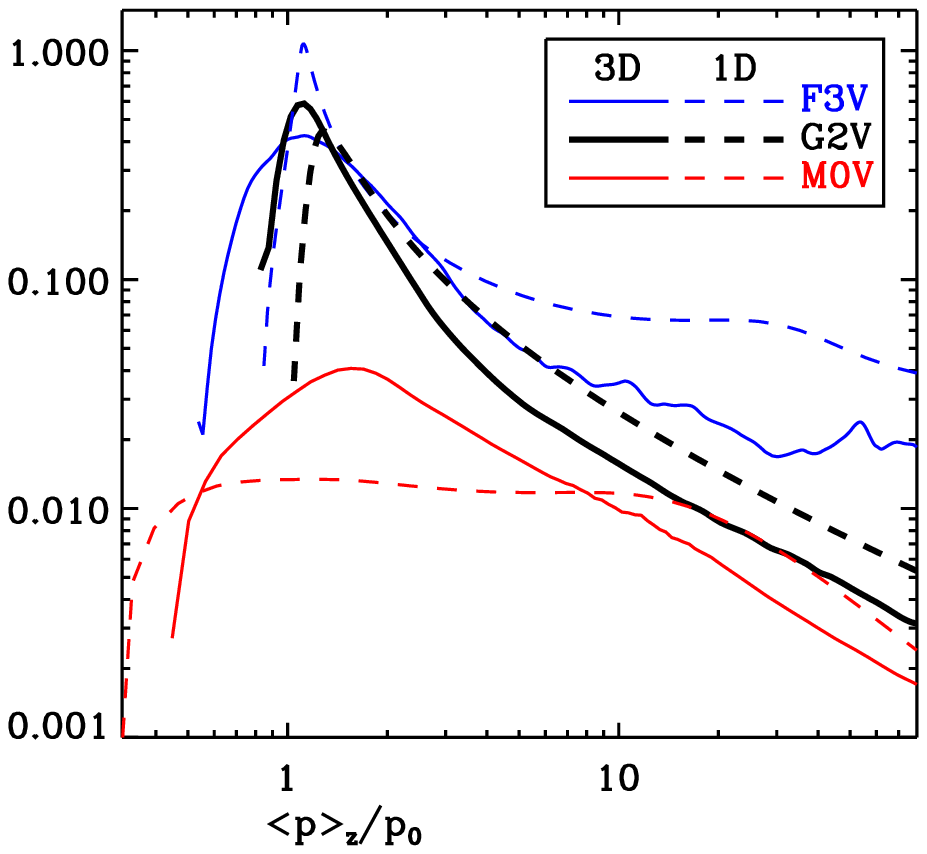}\\
\includegraphics[width=2.95cm]{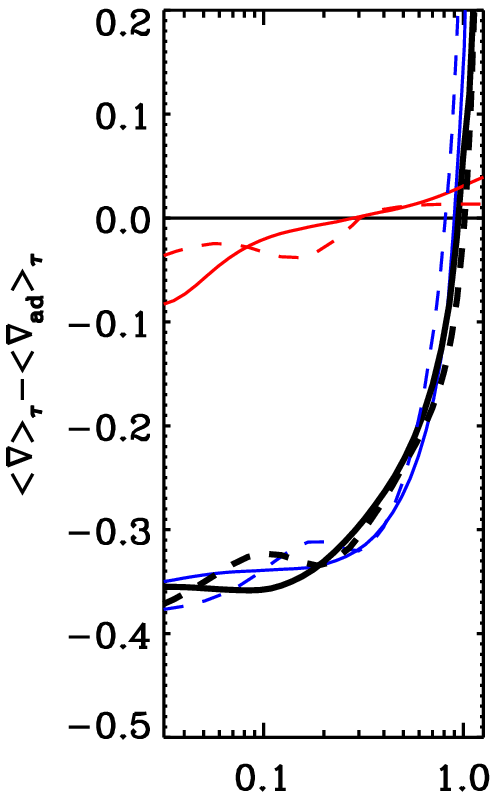}\includegraphics[width=5.55cm]{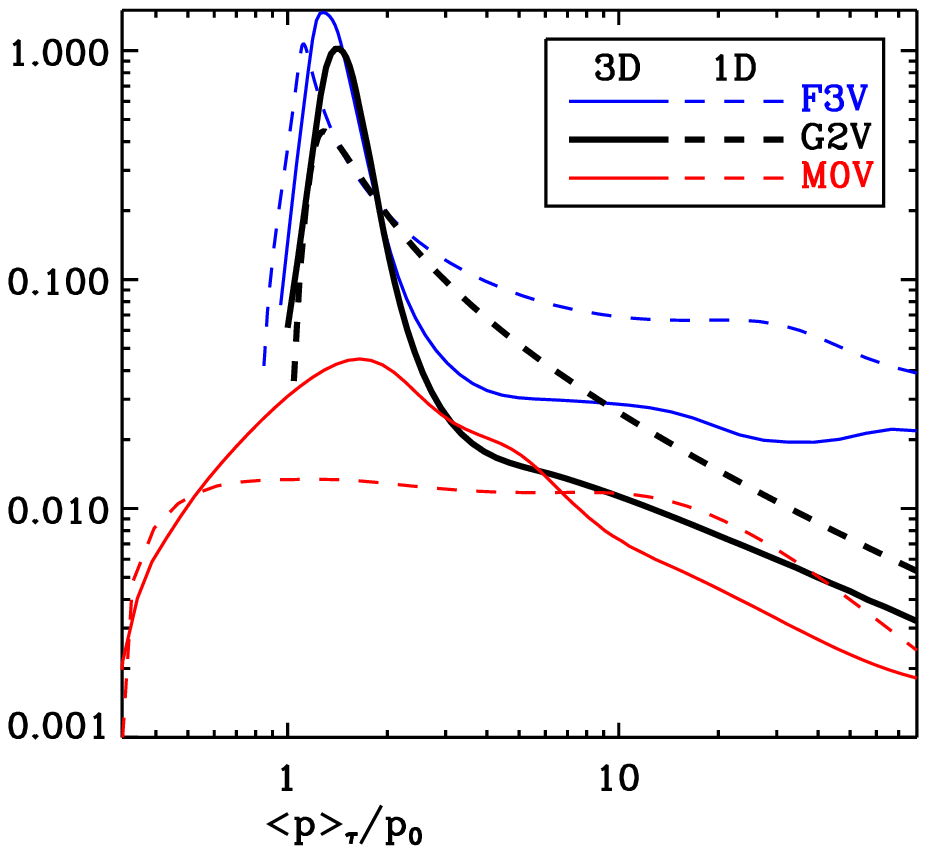}~%
\caption{Comparison of the superadiabaticity of the 3D \texttt{MURaM} simulations with 1D MLT models. {\it Top:} 3D results averaged on planes of constant geometrical depth, $z$. {\it Botom:} 3D results averaged on iso-$\tau_{\mathrm{R}}$ surfaces. The {\it left} subplots show the superadiabaticity of the atmospheric layers on a linear scale, the {\it right} subplots show the superadiabatic regime on a logarithmic scale. In the 1D models, the mixing-length parameter $\alpha$ was set to 1.5, 1.7, and 2.0 for F3V, G2V, and M0V, respectively.}\label{fig:1Dvs3D2}
\end{figure}

In Figure~\ref{fig:1Dvs3D2} the run of the superadiabaticity is shown. The 3D results are shown as iso-$z$ and iso-$\tau_R$ horizontal averages. As discussed in Appendix~\ref{app:avg}, near the photospheric transition, the iso-$\tau$ average is closer to the 1D description and therefore more useful for the comparison between 1D and averaged 3D models although the plain horizontal average is the physically more decisive quantity. Despite the necessary parameterisation of important physics in the 1D models, there is a qualitative similarity between 1D and 3D results. Below its peak, the superadiabaticity of the 1D models is higher by a factor of about 1.5 to 3 compared to the 3D models. For the F3V star, this deviation is the main cause for the difference of the temperature profiles, while for the simulations of the cooler stars, the differences in $\nabla_{\mathrm{ad}}$ related to the equation of state between 1D and 3D models are larger than the small deviation in $\nabla-\nabla_{\mathrm{ad}}$. Around the superadiabatic peak, the
     profiles of 1D and 3D results differ more strongly. In this regime, which is more extended in terms of pressure scale heights for the M0V star than for the other two stars, the superadiabaticity in the 3D models is higher than predicted by the 1D models. The superadiabaticity in the atmospheric layers is qualitatively in agreement between 1D and 3D results.\par
Although there were small deviations in chemical abundances, equation of state, and opacities between 1D and 3D calculations, most of the differences in the upper part of the depth range shown can be attributed to the necessarily very crude treatment of convection and -- most importantly -- radiation in the 1D models versus the comprehensive 3D simulations.\\
\section{Concluding remarks}

We analysed three-dimensional time-dependent hydrodynamical simulations of near-surface convection of six main-sequence stars of spectral types F3 to M2, including one with solar parameters. The coverage in depth (six to eight pressure scale heights below and above the optical surface) allows convective processes to be studied in the upper part of the convective envelope as well as the 3D structure of the atmospheres of the simulated stars.\par
In this first paper of the series, we present the averaged 3D stratification of the six simulations. Moreover, we analysed the energy transport through the upper part of the convective envelope.\par
Although the basic mechanisms driving and shaping the convective processes are similar in all models, the variation of stellar parameters (a factor of 5 in $g$ and a factor of 2 in $T_{\mathrm{eff}}$) entails differences in granule size and shape, intensity contrast, typical flow velocities, temperature fluctuations, scale heights etc. While some of these trends are more or less obvious effects of the variation of the stellar parameters, some are more subtle and are a consequence mainly of the highly non-linear temperature dependence of the opacity. Most notably, the low temperature sensitivity of opacity between 4000 and 5000\,K, which is the main reason for ``hidden'' granulation in simulations of stars cooler than the Sun (especially of K stars), entails some differences in the atmospheric structure between the hotter two and the cooler four simulations of our sequence (see discussions of Figs.~\ref{fig:v},~\ref{fig:strat5},~and~\ref{fig:strat6}). \par 

The differences in the granulation necessarily have an impact on the formation of spectral lines, since line-of-sight velocity, pressure, and temperature (and thus also optical depth) are highly correlated with each other. The details of this impact will be analysed in the second paper of this series where we concentrate on the granulation, spectral line formation, and the 3D structure of the atmosphere.

\begin{acknowledgements}
The authors thank H.-G. Ludwig for providing 1D mixing-length models. They acknowledge research funding by the {\it Deutsche Forschungsgemeinschaft (DFG)} under the grant {\it SFB 963/1, project A16}. BB acknowledges financial support by the {\it International Max Planck Research School }(IMPRS) {\it on Physical Processes in the Solar System and Beyond at the Universities of Braunschweig and G\"ottingen}. AR acknowledges research funding from {\it DFG} grant {\it RE 1664/9-1}.  
\end{acknowledgements}
\bibliography{beecketal}
\appendix
\section{Horizontal averages}\label{app:avg}
\begin{figure*}
\centering
\includegraphics[width=8.5cm]{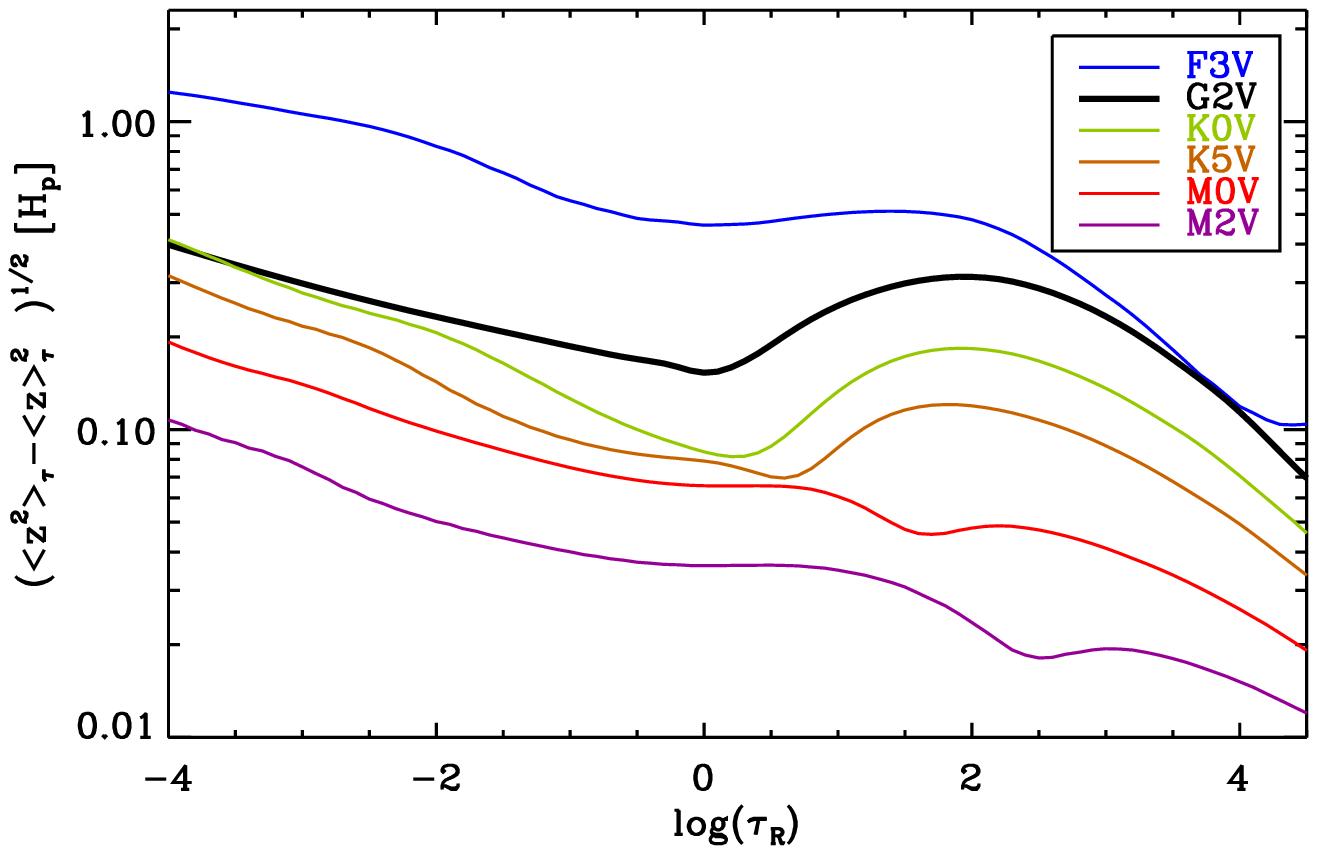}~%
\includegraphics[width=8.5cm]{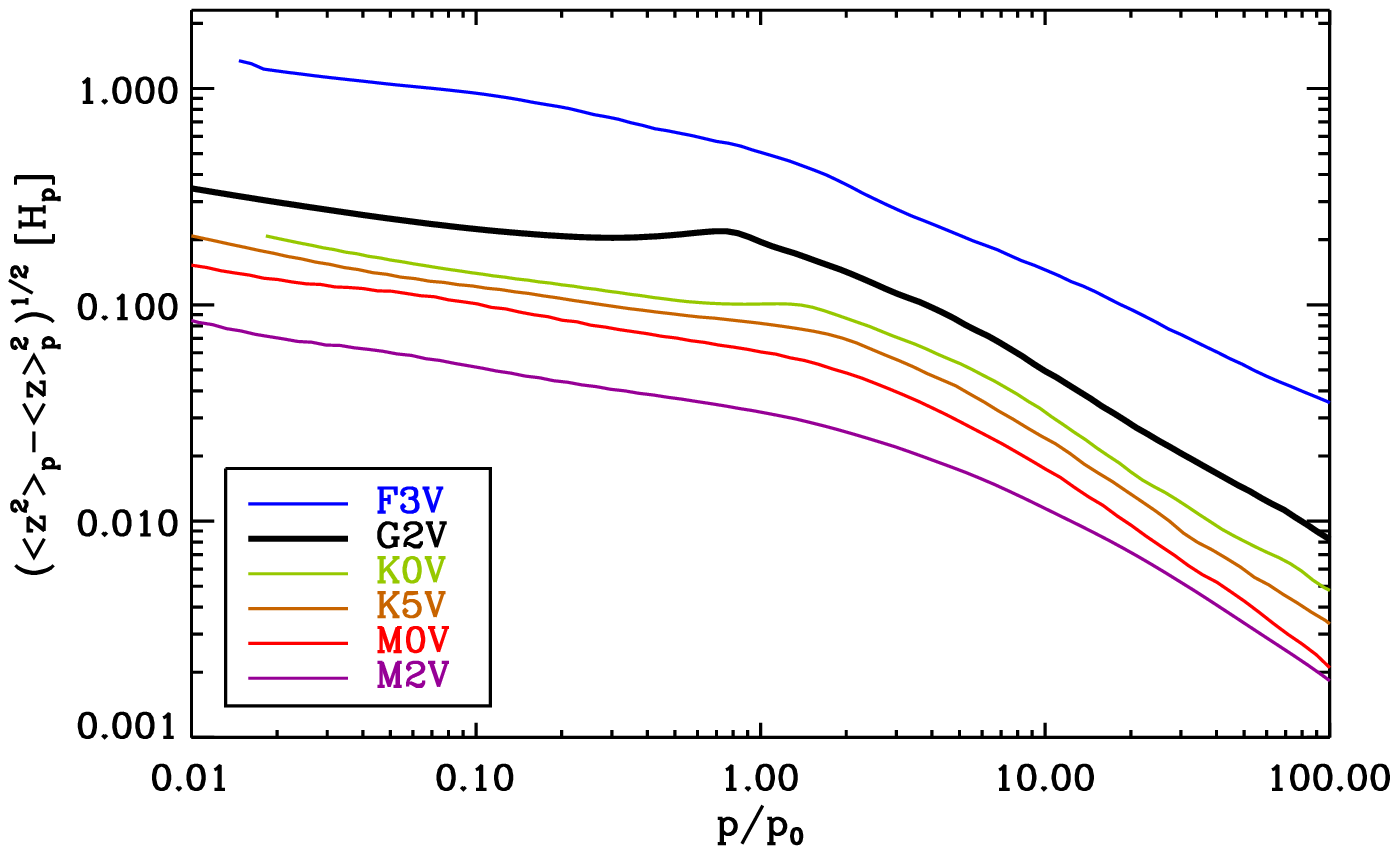}
\caption{Average profiles of the rms fluctuations of the geometrical height on iso-$\tau_{\mathrm{R}}$ surfaces ({\it left}) and on iso-$p$ surfaces ({\it right}) in units of the local pressure scale heights.}\label{fig:corrugation}
\end{figure*}
\begin{figure*}
\centering
\includegraphics[width=8.5cm]{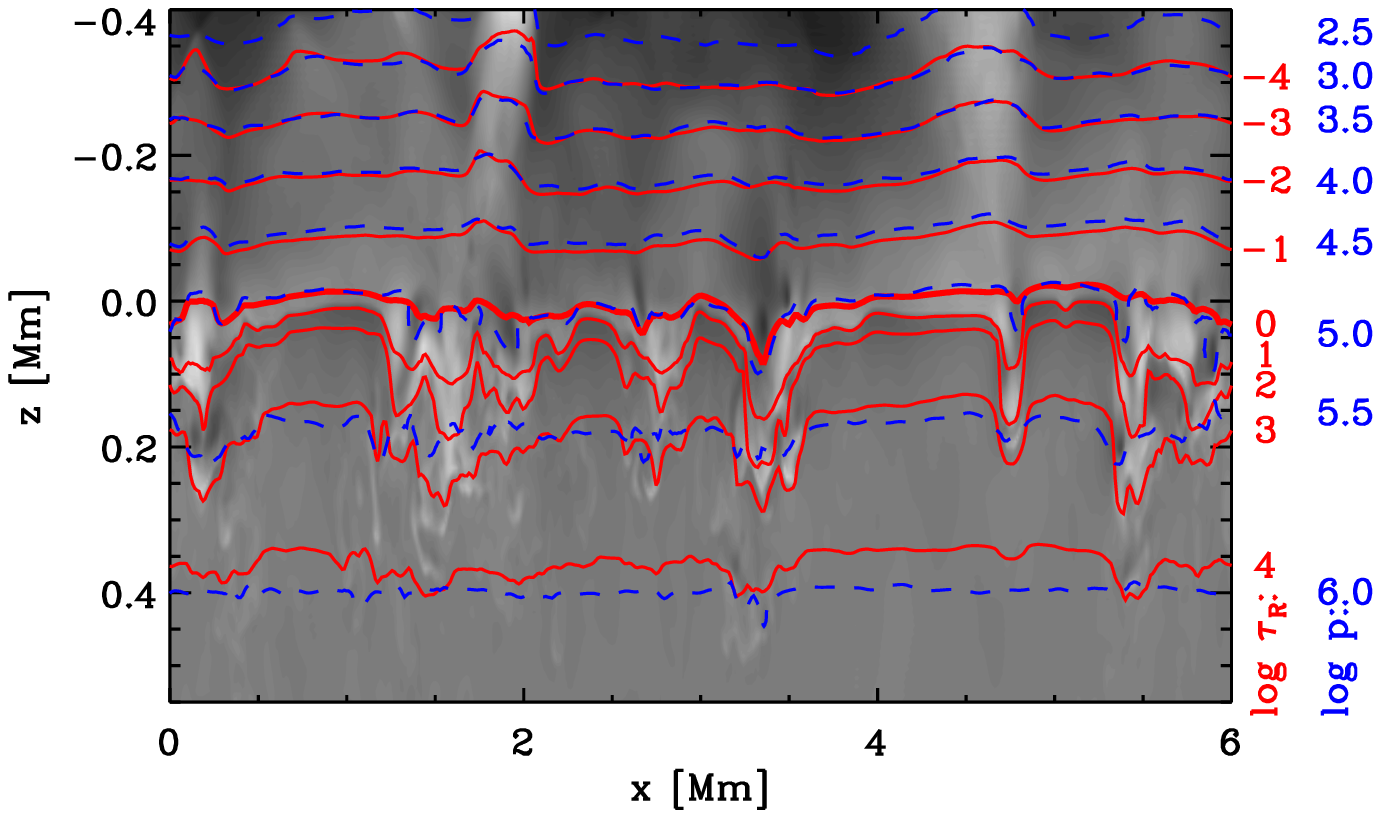}~%
\includegraphics[width=8.5cm]{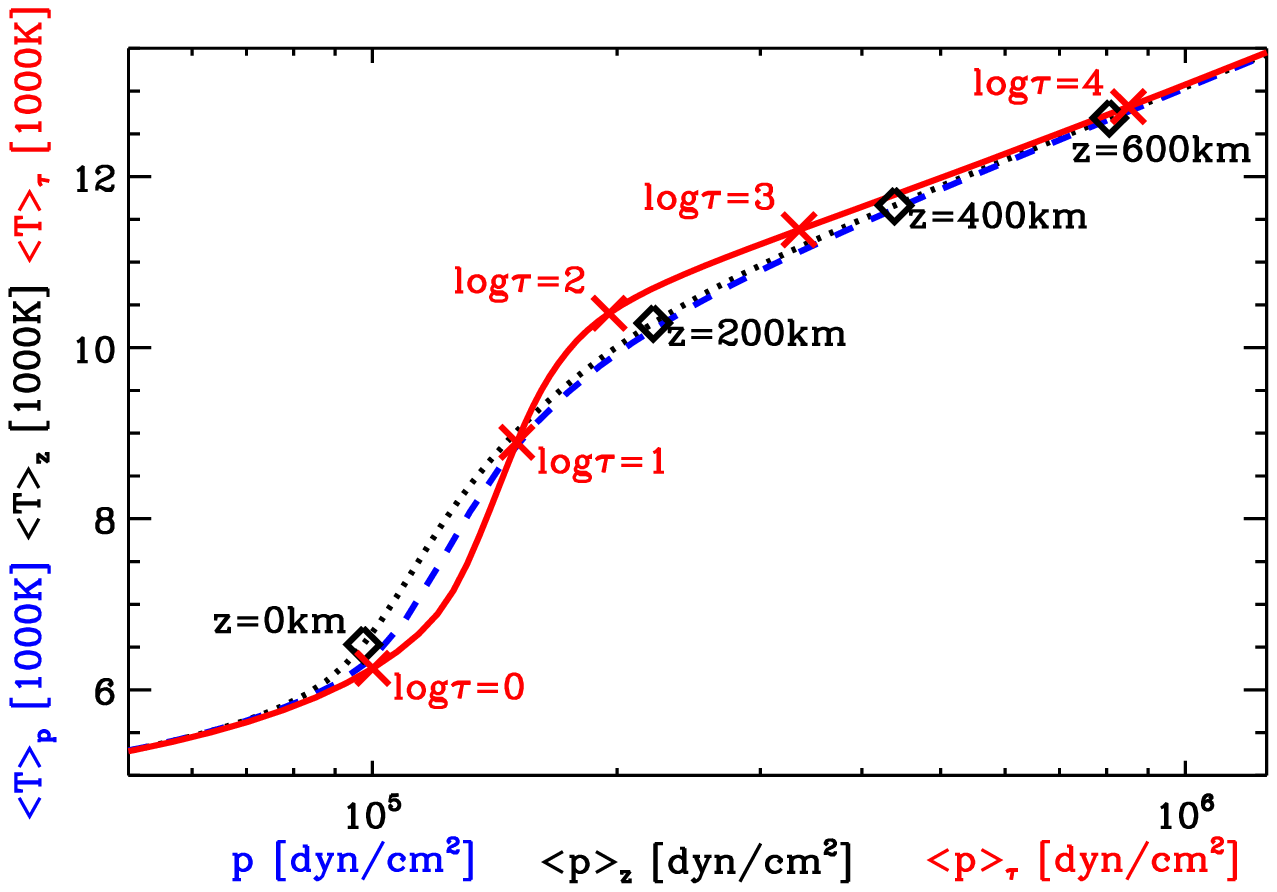}%
\caption{Illustration of the different averages. {\it Left panel:} Vertical cut through a part of the G2V simulation domain. Surfaces of constant optical depth (solid, red curves) and constant pressure (dashed, blue curves) are indicated; the underlying grey-scale image illustrate the density (normalised by the horizontal mean $\langle \varrho\rangle_z$). {\it Right panel:} Run of the horizontally averaged temperature vs. averaged pressure in the G2V simulation for the different averages $\langle\cdot\rangle_z$ (dotted, black), $\langle\cdot\rangle_{\tau}$ (red,solid), and $\langle\cdot\rangle_p$ (blue, dashed); as a reference the position of some $z$ and $\tau$ values is marked.} \label{fig:example}
\end{figure*}
We consider temporally and horizontally averaged quantities to study the mean stratification of the six simulated stars and compare them to 1D models. Depending on the context, the sensible ``horizontal'' average is an average over surfaces of constant geometrical depth, (Rosseland) optical depth, or (gas) pressure, denoted by $\langle \cdot \rangle_{z}$, $\langle \cdot \rangle_{\tau}$, or $\langle \cdot \rangle_{p}$, respectively. The average $\langle\cdot\rangle_{\tau}$ is sensible for the layers around the optical surface and in the photosphere (especially for quantities involving the emergent intensity), while the average $\langle\cdot\rangle_{z}$ is more relevant for the convective, deeper layers, where the stratifications are nearly adiabatic and largely independent from the radiation field. However, $z$ is highly impractical as a coordinate scale for comparing simulations of stars of various spectral types, since the pressure scale height varies by more than a factor of ten between the six simulated stars presented in this paper (cf. Fig.~\ref{fig:scaleheights}). Therefore, we use the logarithm of the normalised pressure $\langle p\rangle_{z}/p_0$ as a more suitable depth coordinate (where $p_0=\langle p \rangle_{\tau_{\mathrm{R}}=1}$ is the average gas pressure at the optical surface) to illustrate depth dependences of quantities averaged on iso-$z$ surfaces. This should not be confused with an average on surfaces of constant pressure, $\langle\cdot\rangle_p$.\par
The averages $\langle\cdot\rangle_{\tau}$ and $\langle\cdot\rangle_p$
are defined as horizontal averages of a quantity on iso-$\tau$
and iso-$p$ surfaces, respectively. For the average $\langle\cdot\rangle_p$, which is rarely used in this article, there is the problem that sometimes iso-$p$ surfaces cannot be unambiguously defined since strong deviations from hydrostatic equilibrium in regions with Mach number of order unity can lead to a locally non-monotonic depth dependence of the gas pressure. Our iso-$p$ surfaces are the deepest surfaces on which the pressure assumes the given value. This arbitrary choice does not significantly influence the results for the iso-$p$ means in the simulations, except for the surface layers of the F3V simulation where the flows are mostly sonic and supersonic (cf. Fig.~\ref{fig:v}).\par
Figure~\ref{fig:corrugation} shows the rms fluctuations of the geometrical depth on surfaces of constant optical depth (left panel) and pressure (right panel) as a measure of the corrugation of these surfaces. Due to this corrugation of the iso-$p$ and iso-$\tau_{\mathrm{R}}$ surfaces, profiles of quantities averaged in the different ways described above are not just distorted versions of each other, but can show a significantly different depth dependence of the same quantity. The differences between the three averaging methods are expected to be largest in the F- and G-type simulations, for which the corrugation of the iso-$p$ and iso-$\tau_{\mathrm{R}}$ surfaces is strongest.\par
The left panel of Figure~\ref{fig:example} shows a vertical cut through some of the iso-$p$ and iso-$\tau_{\mathrm{R}}$ surfaces in the G2V simulation. Note that in the optically thin upper part of the simulation domain, the iso-$p$-surfaces follow the iso-$\tau_{\mathrm{R}}$ surfaces. Although this is observed in all six simulations, this effect is most prominent in the G-type star (see discussion of Fig.~\ref{fig:strat5} in Sect.~\ref{sec:strat}). In the optically thick part, the temperature fluctuations determine the shape of the iso-$\tau_{\mathrm{R}}$ surfaces, since the opacity is highly temperature-sensitive in this regime. The iso-$p$ surfaces, however, become almost flat planes in the deeper layers since the density contrast and deviations from hydrostatic equilibrium decrease with increasing depth.\par
The right panel of Figure~\ref{fig:example} illustrates the different depth dependences of temperature, $T$, for the three different averages $\langle\cdot\rangle_z$, $\langle\cdot\rangle_{\tau}$, and $\langle\cdot\rangle_p$. As for most of the figures in this paper, the gas pressure (here without normalisation) was used as depth coordinate. For $\langle T\rangle_{\tau}$ and $\langle T\rangle_{z}$, the pressure varies along the surfaces over which the average is performed. The depth coordinates are therefore averages themselves, namely $\langle p\rangle_{\tau}$ and $\langle p\rangle_z$, respectively, in these cases. As expected, the differences between the differently averaged temperature are largest at the optical surface and all three averages converge at large optical depth ($\log\tau_{\mathrm{R}} \gtrsim 4$). As the deviations from hydrostatic equilibrium are small in the subsurface layers, $\langle\cdot\rangle_p$ stays close to $\langle\cdot\rangle_z$. The average $\langle\cdot\rangle_\tau$ deviates more strongly near the photospheric transition since the big temperature fluctuations govern the opacity and thus lead to strongly corrugated iso-tau surfaces. In the atmosphere, the deviations between different averages of temperature become smaller with height. Especially, $\langle T\rangle_p\approx\langle T\rangle_{\tau}$, in these layers as expected, because the iso-$\tau$ surfaces roughly follow the iso-$p$ surfaces.\par
If one aims at comparing 1D and averaged 3D results, one has to take into account that a 1D model does not have corrugated iso-$\tau$ surfaces. Profiles of quantities which change rapidly at the photospheric transition have a steep gradient in 1D models comparable to the local gradient in a 3D simulation. As the depth of the photospheric transition varies across the surface in a 3D simulation, these strong local gradients are smeared out and the similarity between 1D and averaged 3D results is obscured, if a plain horizontal average, $\langle\cdot\rangle_z$ is used. The average $\langle\cdot\rangle_{\tau}$ is more appropriate in these cases for a comparison between 1D and averaged 3D profiles. This average however has no relevance below the photospheric transition, where $\langle\cdot\rangle_z$ is more useful. For the stellar parameters used in our simulations, the $\langle\cdot\rangle_p$ average seems a good compromise between the two other averaging methods for comparison with 1D models as it is converging towards $\langle\cdot\rangle_{\tau}$ in the atmosphere and towards $\langle\cdot\rangle_z$ in the convection zone.\par
In order to obtain the temporally averaged profiles presented in this paper, first one of the horizontal averaging methods described above was applied to several snapshots with a time separation of 5 -- 7 minutes, depending on the star. Then, for each quantity and at each depth point, the values of the different snapshots were averaged. The time dependence of the horizontal averages of most quantities under consideration (such as $T$, $p$, $\varrho$, etc.) was found to be very small, so that we found a small number of snapshots to be sufficient for a sensible temporal average. The mean profiles presented in this article are averages over six snapshots each.

\end{document}